\documentclass[twocolumn]{aastex701}

\usepackage{xcolor}
\usepackage{svg}
\usepackage{subfigure}
\usepackage{tabularx}
\usepackage{CJK}        

\newcommand {\libyt}        {\texttt{libyt}}
\newcommand {\yt}           {\texttt{yt}}
\newcommand {\enzo}         {\texttt{Enzo}}

\newcommand {\gamer}        {\texttt{GAMER}}

\newcommand {\km}           {\,\rm km}
\newcommand {\pc}           {\,\rm pc}
\newcommand {\kpc}          {\,\rm kpc}
\newcommand {\Mpc}          {\,\rm Mpc}

\newcommand {\Myr}          {\,\rm Myr}
\newcommand {\Gyr}          {\,\rm Gyr}
\newcommand {\Msun}         {\,\rm{M}_{\odot}}
\newcommand {\ms}           {\,\rm ms}
\newcommand {\MB}           {\,\rm MB}
\newcommand {\GB}           {\,\rm GB}
\newcommand {\eV}           {\,\rm eV}
\newcommand {\MeV}          {\,\rm MeV}

\newcommand {\sref}[1]      {Section~\ref{#1}}
\newcommand {\fref}[1]      {Fig.~\ref{#1}}
\newcommand {\tref}[1]      {Table~\ref{#1}}

\shorttitle{libyt}
\shortauthors{Tsai et al.}

\begin{document}
\begin{CJK*}{UTF8}{bkai}

\title{libyt: an In Situ Interface Connecting Simulations with yt, Python, and Jupyter Workflows}

\author[0000-0003-4635-6259]{Shin-Rong Tsai (蔡欣蓉)}
\affiliation{School of Information Sciences, University of Illinois at Urbana-Champaign, Champaign and Urbana, Illinois, USA}
\affiliation{Department of Physics, National Taiwan University, Taipei, Taiwan}
\email[show]{srtsai@illinois.edu}

\author[0000-0002-1249-279X]{Hsi-Yu Schive (薛熙于)}
\affiliation{Department of Physics, National Taiwan University, Taipei, Taiwan}
\affiliation{Institute of Astrophysics, National Taiwan University, Taipei, Taiwan}
\affiliation{Center for Theoretical Physics, National Taiwan University, Taipei, Taiwan}
\affiliation{Physics Division, National Center for Theoretical Sciences, Taipei, Taiwan}
\email[show]{hyschive@phys.ntu.edu.tw}

\author[0000-0002-5294-0198]{Matthew J. Turk}
\affiliation{School of Information Sciences, University of Illinois at Urbana-Champaign, Champaign and Urbana, Illinois, USA}
\affiliation{Department of Astronomy, University of Illinois at Urbana-Champaign, Champaign and Urbana, Illinois, USA}
\affiliation{National Center for Supercomputing Applications, University of Illinois at Urbana-Champaign, Champaign and Urbana, Illinois, USA}
\email[show]{mjturk@illinois.edu}

\correspondingauthor{Shin-Rong Tsai}

\begin{abstract} 

In the exascale computing era, handling and analyzing massive datasets have become extremely challenging.
In situ analysis, which processes data during simulation runtime and bypasses costly intermediate {disk input and output} steps, offers a promising solution.
We present \texttt{libyt} (\url{https://github.com/yt-project/libyt}), an open-source C library that enables astrophysical simulations to analyze and visualize data in parallel computation with \texttt{yt} or other Python packages.
\texttt{libyt} can invoke Python routines automatically or provide interactive entry points via a Python prompt or a Jupyter Notebook.
It requires minimal intervention in researchers' workflow, allowing users to reuse job submission scripts and Python routines.
We describe \texttt{libyt}'s architecture for parallel computing in high-performance computing environments, including its bidirectional connection between simulation codes and Python, and its integration into the Jupyter ecosystem.
We detail its methods for reading {patch-based adaptive mesh refinement (AMR)} simulations and handling in-memory data with minimal overhead, and procedures for yielding data when requested by Python.
We describe how \texttt{libyt} maps simulation data to \texttt{yt} frontends, allowing post-processing scripts to be converted into in situ analysis with just two lines of change.
We document \texttt{libyt}'s API and demonstrate its integration into two astrophysical simulation codes, \texttt{GAMER} and \texttt{Enzo}, using examples including core-collapse supernovae, isolated dwarf galaxies, fuzzy dark matter, the Sod shock tube test, Kelvin-Helmholtz instability, and the AGORA galaxy simulation.
Finally, we discuss \texttt{libyt}'s performance, limitations related to data redistribution, extensibility, architecture, and comparisons with traditional post-processing approaches.

\end{abstract}


\section{Introduction} \label{sec:introduction}
Scientific simulations have grown exponentially in both size and complexity. Software such as ParaView \citep{ParaView}, VisIt \citep{HPV:VisIt},  MayaVi \citep{ramachandran2011mayavi}, MATLAB, and many others have been widely used in analyzing and visualizing datasets output from simulations. 
More focused, purpose-built tools like \yt\footnote{\yt\ is an open-source, permissively-licensed Python package for analyzing and visualizing volumetric data.} \citep[][\url{https://yt-project.org/}]{yt}, \texttt{pynbody}\footnote{\texttt{pynbody} is an analysis framework for N-body and hydrodynamic astrophysical simulations.} \citep[][\url{https://github.com/pynbody/pynbody}]{pynbody}, and \texttt{tangos}\footnote{\texttt{tangos} is a Python framework and web interface for database-driven analysis of numerical structure formation simulations.} \citep[][\url{https://github.com/pynbody/tangos}]{tangos} have also been developed to suit the needs of quantitative analysis for astrophysical data. 
Even though we can use these tools to accomplish everyday tasks successfully, difficulties arise from having to store the full dataset on disk before performing any further analysis. 
In situ analysis, which analyzes simulation data without the intermediate step of writing data to disk, avoids the difficulties of data storage by directly accessing and analyzing the data within the running simulation.
Furthermore, these tools can reduce the necessary output data size by processing and selecting raw data first.
These improvements are particularly accentuated when only a small fraction of data is needed, when the dataset series in demand has high temporal and spatial resolution that needs ample disk space, or when these factors combine.

Several in situ analysis tools offer a workflow that streamlines the integration of core analysis software and simulations.
The In Situ Analysis Framework, Catalyst \citep{Catalyst}, provides a framework for simulations to link, build upon, and embed visualization using ParaView within the code.
ParaView has a built-in Python client that enables users to write and execute their own Python scripts, thereby automating the data analysis pipeline. 
While extremely capable, the Python client is isolated and designed for ParaView itself, oriented at ParaView operation macros.
The current state of this system is not as flexible or tailored to astrophysical needs, and usage requires an investment in the overall ParaView ecosystem, which is not commonly found in existing simulation codes.
Another tool, Ascent \citep{Ascent}, built explicitly for in situ analysis, is also designed to perform visualization, queries, and data preparation for other software. 
However, users must have prior knowledge of how to program this in situ workflow.
Multiple types of backend services can perform in situ analysis.
SENSEI \citep{SENSEI} provides a generic interface for different in situ analysis backends, 
so that users can easily switch the in situ analysis software mentioned above and many others during runtime. 
It also provides an interface to run Python code, but requires users to do it in a specific format.
Other relevant libraries offer methods for binding C/C++ and Python applications without directly mapping memory structures to analysis-ready objects.
For example, \texttt{pybind11} \citep{pybind11} provides methods to expose C/C++ objects to Python and vice versa, and support methods to run Python code inside the C/C++ applications. 
Still, it is not seamlessly integrated into the entire simulation and data processing workflow,
which requires users to implement everything from scratch.

{Beyond these established frameworks, recent efforts have further expanded the in situ and high-performance data analysis ecosystem.
This includes improving the data transport layers that enable in situ analysis, such as ADIOS2 \citep{ADIOS2}, Wilkins \citep{Wilkins}, DEISA \citep{deisa}, and PDI (\url{https://github.com/pdidev/pdi}).
The in situ technique has also been applied in fluid dynamics applications \citep{ImpactOfInSituTechniques, InSituTechniquesOnGPU}, cosmological simulations \citep{tuccari2025exascaleinsituvisualizationastronomy}, as well as hydrodynamics simulations \citep{yan2025realtimeautoregressionmethodinsitu}, which demonstrate its capability in solving challenges for data storage.
}
{The tools described above can mitigate the expensive disk input and output before extracting scientifically meaningful results and reduce the technical burden of integrating analysis routines into simulation codes. Some also provide Python interfaces for constructing analysis pipelines. However, we still found a practical gap between the researchers and these tools.
What remains missing is a lightweight and flexible in situ library that integrates naturally with researchers' everyday Python workflows and environments and can incorporate arbitrary Python scripts with minimal effort.}

While simulations are typically implemented in high-performance languages such as C, C++, Fortran, or other emerging languages, many data analysis tools and data-driven workflows, such as machine learning frameworks and \yt, are written in Python. 
Although these codes typically are backed by numerical code written in lower-level languages (such as C, or in some cases ``just-in-time'' compiled languages), the interface is presented in a higher-level abstraction.
Thus, being able to interchange Python code for post-processing and for in situ analysis provides a reduced barrier to entry and greater reuse of analysis tools and scripts.
Moreover, some machine learning frameworks require specialized Python setups.
Executing the in situ analysis Python script within the same environment used for daily Python tasks reduces the risk of calling incorrect dependencies and unexpected routines during runtime.

To facilitate this, we developed \libyt\ {\citep[][\url{https://github.com/yt-project/libyt}]{libyt-zenodo}}, 
a C library for \yt, which does in situ analysis in a user-designated Python environment by directly calling \yt. 
A fundamental design goal of \libyt\ is to provide an interface that mirrors the experience of using \yt\ for post-processing simulation data on disk, allowing a post-processing script to be converted to an in situ script with only a two-line change.
This reusability of post-processing scripts, combined with the ability to run in situ analysis within an already configured Python environment, significantly reduces the learning curve.
We can monitor simulation progress in real time and run arbitrary Python scripts to process distributed in-memory simulation data.
This includes extracting a subset from the full dataset, analyzing a time series of data with high temporal resolution, and even exploring data using the Jupyter Notebook user interface or interactive Python prompt.
Additionally, \libyt\ enables back-communication of simulation information, allowing researchers to define functions that generate derived fields requested by Python. 
This mechanism enables the direct reuse of existing routines in the simulation code (for instance, those used to compute the equation of state).
{By avoiding disk read and write operations, \libyt\ minimizes the overhead associated with disk access during data analysis, albeit at the cost of some additional memory consumption.
Analysis can be conducted with much greater frequency}, providing more fine-grained insight into the simulation's progress.
\libyt\ is not limited to local machines; it also works on high-performance computing clusters, which greatly simplifies debugging simulations when scaling up in cluster environments.
Overall, \libyt\ tightly integrates simulations with the Python ecosystem and enables human-in-the-loop interaction via Jupyter and interactive prompts without sacrificing their usability.


\libyt\ has its origins in the \enzo\  project \citep[][\url{https://github.com/enzo-project/enzo-dev}]{enzo}. \enzo\ is a community-developed adaptive mesh refinement (AMR) simulation code, designed for rich, multi-physics hydrodynamic astrophysical calculations. \enzo\ supports in situ analysis with \yt, but this feature is relatively deeply embedded and is not easily extracted.  Also, the functionalities that \yt\ supports are limited and are not robust against errors or exceptions. 
We extract the embedded Python feature from \enzo\ into \libyt\ and then greatly extend its efficacy. 
This C library can now be applied to other simulation codes and supports most \yt\ functions.
\libyt\ was first introduced as a C++ library to perform in situ Python analysis within a running simulation, and it was subsequently augmented with an interactive Python prompt feature \citep{libyt-scipy2023}.
Next, \libyt\ was updated to become a pure C library, and we added a new extension that supports interactive access to running Jupyter kernels launched by the simulation \citep{libyt-pasc-2024}.
This paper aims to provide a detailed description of the architecture of \libyt, demonstrate the usage of the \libyt\ API, and present simulation results and applications from several astrophysical production runs.


We start by describing the software architecture, the technical details, and how \libyt\ is built in Section \ref{sec:code-method}. 
We then walk through the procedure of implementing \libyt\ into simulations in Section \ref{sec:demonstration}. 
We demonstrate some production runs
with the GPU-accelerated AMR code \gamer\ \citep[][\url{https://github.com/gamer-project/gamer}]{gamer-2} and \enzo\ 
using \libyt\ and show that it plays an important role in large simulations in Section \ref{sec:applications}.
We discuss \libyt's performance, limitations, comparison to the traditional post-processing method, and potential use case of binding to AI and machine learning in \sref{sec:discussion}, and we conclude in \sref{sec:conclusions}.

\section{Software Architecture} \label{sec:code-method}

This section describes the methods and techniques to build \libyt\ and its related components.
We first give an overview of the architecture of \libyt\ and related libraries built to connect to other services in \sref{subsec:overview}.
We then break into four parts: binding simulation objects to Python and enabling back-communication from Python to simulations (\sref{subsec:embedding-python-in-c++-application}); running Python codes and providing entry points for interactive exploration of data (Section \ref{subsec:executing-python-codes-and-python-prompt} --- \ref{subsec:support-jupyter-notebook}); how we define the data structure that aligns with \yt\ and make reusing post-processing scripts achievable (Section \ref{subsec:connecting-libyt-to-yt} --- \ref{subsec:amr}); and, how to deal with data redistribution when doing in situ analysis (\sref{subsec:parallelism}).

\subsection{Architecture of Libyt} \label{subsec:overview}

\begin{figure*}[ht]
\begin{center}
\includegraphics[width=0.8\textwidth]{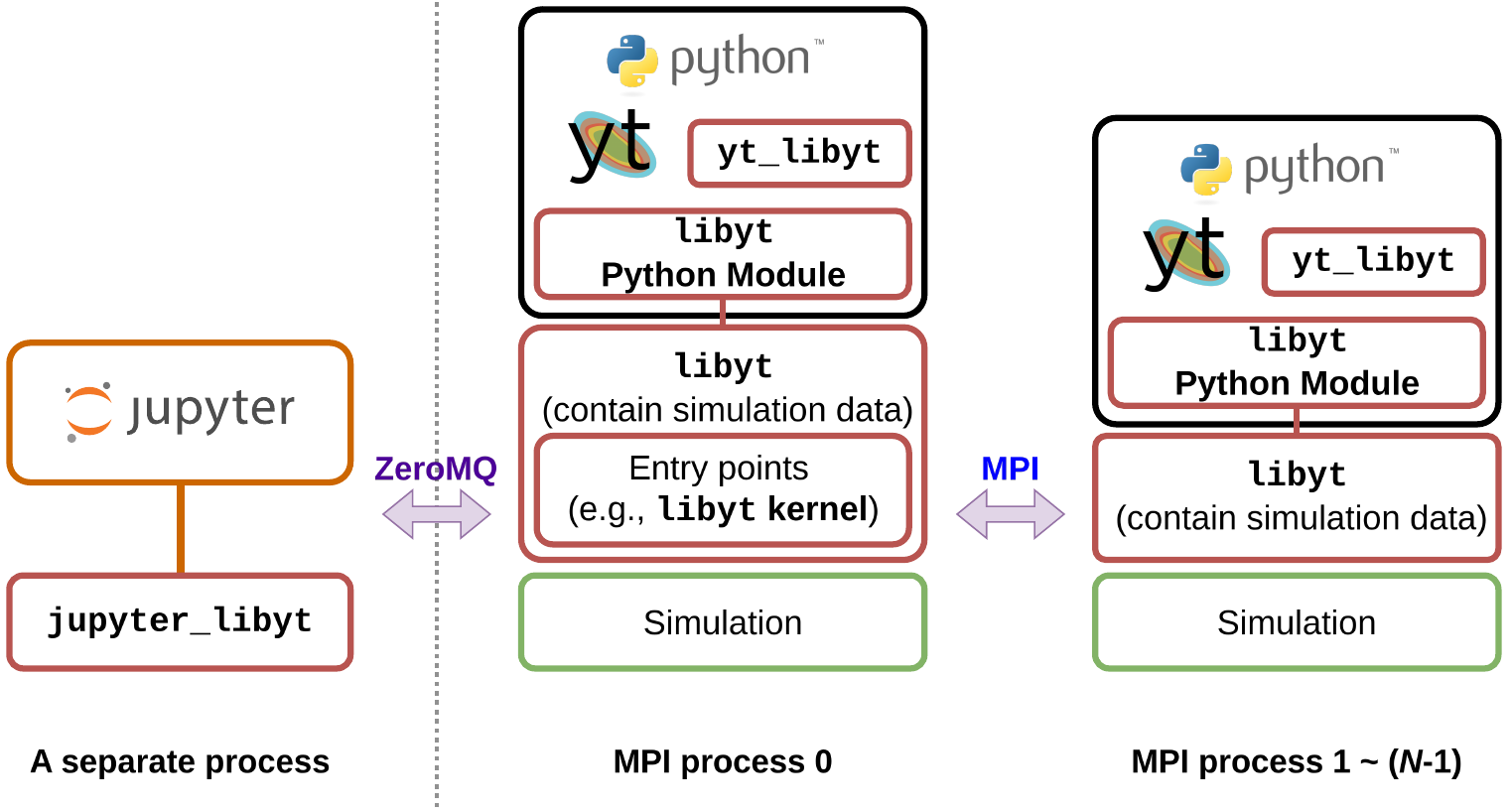}
\end{center}
\caption{ 
The overall structure of \libyt\ and all the related components: \libyt\ Python Module, \libyt\ kernel, \texttt{yt\_libyt}, and \texttt{jupyter\_libyt}, that are built around it to integrate simulations with Python, \yt, and Jupyter.
Although \libyt\ can run arbitrary Python scripts, we focus on its use with \yt.
The library provides an API for binding simulation data and functions to Python, invoking analysis routines, and enabling entry points, such as the \libyt\ kernel for Jupyter Notebook.
During in situ analysis, the original simulation processes pause and wait until Python completes all jobs.
A Jupyter Notebook running in a separate process connects to the \libyt\ kernel via ZeroMQ via the \texttt{jupyter\_libyt} extension, with all Python interpreters and simulations working synchronously under the MPI task space.
The details of each component are described in \sref{subsec:overview}.
}
\label{fig:overview}
\end{figure*}

\libyt\ provides a bridge that connects simulations to Python instances bidirectionally.
Components built around \libyt\ are responsible for integrating other well-developed projects, such as Jupyter and \yt, to minimize the learning curve from the user's perspective.
\fref{fig:overview} illustrates \libyt's overall structure and other related libraries created to connect to other existing software. 
{The purpose and role of each library is outlined below, from simulation-side data binding and control, through Python integration and \yt\ analysis layers, to the Jupyter interfaces that allow users to explore simulation data at runtime}:
\begin{itemize}
\item ``\texttt{libyt}'' provides an API for simulations to bind data and functions, run Python code to perform in situ analysis, and expose entry points such as Python prompt, file-based Python prompt, and Jupyter Notebook for users to interact with data using Python syntax.
\item ``\texttt{libyt} Python Module'' is the hub and the Python counterpart of \libyt. It is a container that stores bound simulation data and objects, and it provides C-extension methods for Python to request data from simulations, enabling back-communication. (\sref{subsec:embedding-python-in-c++-application})
\item ``\texttt{yt\_libyt}'' is a \yt\ frontend for \libyt. A \yt\ frontend abstracts the data structures and information in a simulation, so that \yt\ understands how to process a simulation dataset. (\sref{subsec:connecting-libyt-to-yt})
\item ``\texttt{yt}'' is a Python package for analyzing and visualizing volumetric data, and it is the analytic package used in this paper. \yt\ has been applied in various domains, including astrophysics, seismology, nuclear engineering, molecular dynamics, and oceanography. This paper demonstrates its application in astrophysics. (\sref{sec:applications})
\item ``\texttt{libyt} kernel'' is a Jupyter kernel. It is launched by the \libyt\ API and serves as an entry point for users to interact with simulation data during runtime. (\sref{subsubsec:libyt-kernel})
\item ``\texttt{jupyter\_libyt}'' is a Jupyter extension for the Jupyter Notebook to connect to a \libyt\ kernel. It handles the kernel and configures the user interface in Jupyter Notebook when connecting to a \libyt\ kernel. (\sref{subsubsec:libyt-provisioner})
\end{itemize}

\libyt\ utilizes the Python C API \citep{python3} and the NumPy C API \citep{numpy} to bind simulation metadata and arrays of data to Python.
In addition, \libyt\ provides methods for Python to call user-defined simulation functions, enabling bidirectional communication between Python and the simulation.
\libyt\ Python Module is the anchor for registered simulation data and the proxy for calling user-defined C functions.
See \sref{subsec:embedding-python-in-c++-application} for details.
With this infrastructure, the embedded Python instances can access and request data from simulations, as well as call Python packages to analyze the data.

When conducting large simulations in parallel using the Message Passing Interface (MPI) \citep{mpi31}, simulation data are distributed in multiple processes. 
Thus, it is crucial for \libyt\ to support MPI as well.
When launching a simulation using \textit{N} MPI tasks, 
each process will create and launch a \libyt\ Python Module and an embedded Python interpreter.
These \textit{N} Python instances will work together.
{\libyt\ handles data input and output (data I/O), referring to the data exchange layer between simulations and Python instances as well as data movement across computing nodes, entirely in memory without accessing disk storage.
This includes collecting data from local compute nodes and redistributing them among MPI processes.}
The latter is essential for the \textit{N} Python instances to conduct in situ analysis in parallel,
since the data required by a Python instance may reside in multiple processes.
See \sref{subsec:parallelism} for details.

When doing in situ analysis, all the simulation computing processes stop, 
and \libyt\ runs Python scripts synchronously in the MPI tasks space. 
This is very similar to running Python scripts in post-processing (\sref{subsec:performing-in-situ-analysis} gives an example), except that data are stored in memory instead of disk.
The runtime switches from simulation to embedded Python interpreters.
During this stage, \libyt\ can also expose entry points for users to interact with the Python instances. 
Entry points, such as the interactive Python prompt, file-based Python prompt, and Jupyter Notebook, are supported for users to redefine the in situ process, enter Python statements, and process simulation data at the current simulation timestamp.
The methods for creating the prompt and running Python codes are described in \sref{subsec:executing-python-codes-and-python-prompt}.
The design of the Python prompt and connecting to the Jupyter Notebook user interface are described in \sref{subsec:interactive-mode} and \sref{subsec:support-jupyter-notebook}.

Despite its name, \libyt\ is a general-purpose library that can launch arbitrary Python scripts and Python modules; however, in this paper, we focus on using \yt\ for in situ analysis.
The API for storing AMR data is designed to match the data structure of \yt, 
allowing post-processing script to be used interchangeably for in situ analysis (see \sref{subsec:connecting-libyt-to-yt} and \sref{subsec:amr}).

\subsection{Embedding Python in C/C++ Applications} \label{subsec:embedding-python-in-c++-application}
When embedding Python in a C/C++ application, Python code isn't compiled ahead of time into a binary.
Instead, the application embeds a Python interpreter that executes Python scripts at runtime.
Python provides a C API that enables users to initialize the Python interpreter, build Python modules, create and access Python objects, and execute Python code within a main C/C++ program.
NumPy combines the effective performance of C with the flexibility of Python. 
NumPy's separation of the underlying memory data block and its definition (e.g., stride, data type) allows wrapping simulation data into a NumPy array object and interpreting it in Python without duplication.
NumPy provides a C API for users to create NumPy array objects so that
Python can interpret the memory block allocated by a C or C++ application.
This includes wrapping an existing C/C++ array into a NumPy array or allocating a new one from scratch and then populating it with data.
{\libyt\ uses the Python C API and NumPy C API to manage reference counting and conversions between Python objects and C/C++ types, thereby ensuring seamless data exchange and precise control between simulation codes and Python-based analysis workflows.}

When \libyt\ is first initialized, it dynamically constructs a Python module, called ``\libyt\ Python Module''.
\libyt\ Python Module functions only as a shim as \libyt's Python counterpart, providing a simple conduit that lets Python access simulation data or request it.
It is only importable during runtime, using:
\\
\\
{\small
\begin{tabular*}{\columnwidth}{l}
\toprule
\texttt{import libyt} \\
\hline
\end{tabular*}
}

Each process has its own unique instance of \libyt\ Python Module in memory, which handles data I/O between Python and the simulation.
Python interpreter uses the module to operate on the data.
\libyt\ Python Module
(i) stores registered simulation information in Python dictionary objects, and
(ii) provides Python C-extension methods for Python to request data from simulations,
as listed in \tref{tab:libyt-python-module}.
We will describe how these Python objects and NumPy arrays are constructed in the following
subsections and demonstrate how the data are filled in in \sref{sec:demonstration}.

\begin{deluxetable*}{lll}
\tablecaption{\libyt\ Python Module\label{tab:libyt-python-module}}
\tablehead{
\colhead{Python object type} & \colhead{Python object} & \colhead{Usage}
}
\startdata
    Dictionary & \texttt{libyt\_info} & \libyt\ information, such as version number and build options. \\
    Dictionary & \texttt{grid\_data} & Intrinsic field data, such as gas mass density
         and gravitational potential, \\
    & &  which are stored in a contiguous array. \\
    Dictionary & \texttt{particle\_data} & Intrinsic particle data, such as particle mass and position, that are stored \\ 
    & & in a contiguous array. \\
    Dictionary & \texttt{hierarchy} & AMR grid hierarchy, such as refinement level,
         grid indices, parent grid \\
    & & indices, grid dimension, and grid coordinates (see \sref{subsec:amr}). \\
    Dictionary & \texttt{param\_yt} &  Built-in \yt\ and \libyt\ parameters, such as
         simulation time, domain \\ 
    & & dimensionality, code units, and number of grids. \\
    Dictionary & \texttt{param\_user} & Frontend-specific parameters, such as
         equation of state parameters. \\
    Dictionary & \texttt{interactive\_mode} & It is available only in interactive mode and records information such as error  \\
    & & messages from Python functions and their definitions (see \sref{subsec:executing-python-codes-and-python-prompt}). \\
    \hline
    C-extension method & \texttt{derived\_func} & Get derived field data from the local MPI process. \\
    C-extension method & \texttt{get\_field\_remote} & Get intrinsic or derived field data from other computing nodes. \\
    C-extension method & \texttt{get\_particle} & Collect intrinsic or derived particle data from the local MPI process. \\
    C-extension method & \texttt{get\_particle\_remote} & Collect intrinsic or derived particle data from other computing nodes. \\
\enddata
\tablecomments{\libyt\ Python Module is a lightweight shim for Python to get simulation data. It is only constructed once when \libyt\ is initialized.}
\end{deluxetable*}

All existing data objects are created under nested Python dictionaries categorized by their usage.
When binding small, discrete, single-piece simulation metadata, \libyt\ creates a new Python object under the \libyt\ Python Module dictionary.

When dealing with memory blocks, such as simulation field data or particle data stored in a contiguous array, we do not copy them to minimize memory overhead.
Instead, \libyt\ instructs Python on how to interpret these memory blocks by using the NumPy C API\footnote{\texttt{PyArray\_SimpleNewFromData} creates an array wrapper around the provided data pointer, allowing Python to interpret the underlying memory.} and providing the necessary metadata (e.g., dimensions and data type).
This turns the memory blocks into NumPy arrays without duplicating the data.
Existing memory blocks belong to and are managed by the simulation process.
To prevent them from being altered accidentally, the arrays are marked as read-only\footnote{The writable flag \texttt{NPY\_ARRAY\_WRITABLE} of the NumPy array is cleared using \texttt{PyArray\_CLEARFLAGS}, making the array read-only.}. 
This creates a new read-only NumPy array using the simulation data as the underlying source, allowing Python to access it directly.
The resulting NumPy arrays are stored under \libyt\ Python Module dictionaries.

The above mechanisms are all initiated by the simulation.
To go the other way around and enable the back-communication initiated by Python, we construct \libyt\ Python Module C-extension methods.
The C-extension methods serve as a proxy for user-defined C functions.
When the C-extension method is called from Python, it first allocates new memory blocks, then invokes the user-defined C function while passing pointers to those blocks.
The C function generates and writes data into the allocated memory.
Afterward, the memory blocks are wrapped into NumPy arrays, and ownership of the arrays is transferred to Python\footnote{The own-data flag \texttt{NPY\_ARRAY\_OWNDATA} of the NumPy array is enabled using \texttt{PyArray\_ENABLEFLAGS}, giving Python ownership of the memory.}. 
It is now Python's responsibility to free these arrays when the reference count drops to zero.
Because the data are generated only when requested from Python, these data are \textit{virtual}.

These mechanisms bind the AMR simulation data structure and data (e.g., field and particle data, derived data) to the \libyt\ Python Module, as described in \sref{subsec:amr}.
Once bound, the simulation data become available under \libyt\ Python Module.

\subsection{Python Scripts and Python Shell} \label{subsec:executing-python-codes-and-python-prompt}

\sref{subsubsec:importing-and-calling-python} describes how \libyt\ imports the Python script and calls the in situ analysis workflow wrapped in Python functions during simulation runtime.
Both fail-fast and fault-tolerant Python calls are supported.
Fail-fast Python refers to a mode in which the entire process immediately terminates if any Python error occurs, which could be a syntax error or a runtime exception.
In contrast, fault-tolerant Python allows the process to continue running by capturing and reporting Python errors without shutting down the simulation.

\libyt\ also provides a Python read-evaluate-print loop (REPL), which acts like a standard Python prompt but with simulation data loaded in advance.
\sref{subsubsec:python-prompt-and-libyt-commands} describes how the Python shell orchestrates MPI processes
and how it retrieves and manages Python function information.
It is the base of the interactive and file-based Python prompt (\sref{subsec:interactive-mode}),
and Jupyter Notebook feature (\sref{subsec:support-jupyter-notebook}).

\subsubsection{Importing the Script and Calling In Situ Workflows} \label{subsubsec:importing-and-calling-python}
After initializing the Python instances, each process imports the user's Python script.
The Python script imported will serve as the namespace, and all Python code will run inside this environment. 
This script can define multiple functions or Python callable objects so that we can use \libyt\ API to call them when performing in situ analysis at different simulation stages.
These functions support loading arguments from simulation codes, which is important for flexibility.
Because \libyt\ only imports the script once during initialization and subsequently calls Python functions multiple times for in situ analysis, any operations and variables that only need to be initialized once should be put outside of a function to prevent unnecessary redefinition.
This pre-written script cannot be changed during runtime in fail-fast Python.
The whole simulation process shuts down if an error occurs while running these Python functions, allowing fail-fast in the production run. 

\libyt\ also provides a build option for a fault-tolerant Python runtime.
In this mode, error messages will be stored, and the simulation continues without interruption.
Users can update and redefine Python functions and variables through the entry points, 
like interactive and file-based Python prompt (\sref{subsec:interactive-mode}) and Jupyter Notebook (\sref{subsec:support-jupyter-notebook}). 
Updating the Python functions is only available in fault-tolerant mode.

The in situ workflows are encapsulated inside Python functions. 
There are no constraints on the function name or the number of invocations; any Python callable objects are valid.
When calling the Python function, the simulation processes halt until the Python function completes.
\libyt\ API (see \sref{subsec:performing-in-situ-analysis}) prepares the function name along with any existing arguments, and then uses Python C API to execute the function in the script's namespace\footnote{We wrap the statement in Python \texttt{exec} built-in function and invoke \texttt{PyRun\_SimpleString}. In fault-tolerant mode, the statement is further wrapped inside a \texttt{try}--\texttt{except} block.}.
When operating in fault-tolerant mode, any errors raised during execution are recorded in the \libyt\ Python Module dictionary and can be retrieved afterward.
This operation sounds simple, but it requires careful coordination between Python and {the} simulation (see \sref{subsubsec:python-prompt-and-libyt-commands}).

\subsubsection{Python Shell and Libyt Commands} \label{subsubsec:python-prompt-and-libyt-commands}
The Python shell follows a {REPL} workflow.
It accepts both Python code and \libyt-defined ``magic'' commands as inputs.
{Executing user inputs consistently across all MPI processes presents several challenges, including (i) determining when the user has finished inputting and handling multi-line statements (e.g., \texttt{def}, \texttt{if}, triple quotes, etc), (ii) ensuring synchronized execution across processes, and (iii) collecting output from each process and presenting it to the user.
To address these issues, \libyt\ designates a single MPI process to coordinate the REPL workflow.}

{
During the user-input phase, the designated process is responsible for determining when the user has finished entering code, validating its syntax, and broadcasting the code to all other processes.
In the meantime, all other processes remain blocked until the broadcast is received.
A complete statement, whether single-line or multi-line, is used to indicate that the user has finished input.
}
The designated MPI process first compiles the Python statement into a bytecode object\footnote{Python C API \texttt{Py\_CompileString} compiles strings into a bytecode object.}. 
{If the compilation is successful, it broadcasts the statement to the other processes, which then compile it locally into bytecode objects.
Subsequently, all processes enter the execution phase.
}
{If the compilation fails, the designated MPI process determines whether the error results from incomplete input in a multi-line statement. 
In such cases, it waits for additional user input; otherwise, the code is treated as a syntax error, and the associated error messages are displayed.
For file-based Python prompt (\sref{subsec:interactive-mode}) and Jupyter Notebook (\sref{subsec:support-jupyter-notebook}), which accept multiple statements at once, the code is treated as a syntax error immediately, since the user explicitly submits the input.
}

During the code execution {phase}, all processes run their bytecode objects inside the imported Python script namespace (\sref{subsubsec:importing-and-calling-python}), and \libyt\ temporarily redirects Python's standard output and error streams to in-memory text buffers.
After execution completes, each process collects the captured output and error messages, and \libyt\ gathers them to the designated MPI process.
Python's standard output and error streams are then restored to default settings.
Though the mechanism can orchestrate MPI processes {and ensure synchronized code execution}, running the code and harnessing the output strings loop has the caveat of being unable to deal with interruptions.
We discuss this limitation in \sref{subsec:limitations}.

Every change made will be stored under the imported Python script's namespace, unless it is deleted using Python's built-in function \texttt{del}.
It accumulates new Python objects and overwrites existing ones, giving us the flexibility to rewrite the in situ workflow and redefine Python functions.
We can think of it as running the Python statement line-by-line in a regular Python prompt, which will hold functions, variables, Python objects, etc., throughout the in situ analysis.

\libyt\ provides a set of IPython-style \citep{IPython} magic commands that allow users to check the status, view error messages, and inspect the function definition of each Python function, control whether a function is run in subsequent iterations, export the input history, and load a Python script.
Every \libyt\ defined command starts with \texttt{\%libyt}.
When retrieving the Python function execution status and error messages, \libyt\ checks if an error string is registered in the \libyt\ Python Module dictionary (\sref{subsec:embedding-python-in-c++-application}).
Python function definitions are likewise registered in the \libyt\ Python Module dictionary, which always reflects the most up-to-date version of each function in the namespace\footnote{\libyt\ retrieves function bodies in a script using Python's \texttt{inspect} module and identifies function definitions in input statements by testing for Python callables.}.
\libyt\ records every valid Python statement and \libyt\ command to support history tracking and reproducibility. 
The history cache is cleared once the entry point terminates.
When loading the Python script, it follows the {REPL} workflow described in this section.
\libyt\ runs the input file line-by-line inside the imported Python script namespace.

\subsection{Interactive Mode} \label{subsec:interactive-mode}

\libyt\ has an interactive Python prompt feature, like a typical Python prompt, but with access to running simulation data from \libyt\ Python Module described in \sref{subsec:embedding-python-in-c++-application}.
The prompt is populated on the MPI root process, allowing a single Python statement at a time.
The root process commences with reading, broadcasting instructions or codes, gathering evaluated outputs, and printing overall results.
It first reads user inputs.
Then, it calls the Python shell described in \sref{subsubsec:python-prompt-and-libyt-commands}.
After printing the results in every MPI process, it awaits the user further. 
In the interactive Python prompt, simulation processes are halted indefinitely until the user exits the prompt.
Users must submit interactive jobs (e.g., \texttt{qsub -I} for a PBS scheduler) when using this feature in a high-performance computing cluster.

To overcome this limitation, \libyt\ also provides a file-based Python prompt feature.
The file-based Python prompt has the same features as an interactive Python prompt, except that it avoids entering Python statements through the terminal and instead reads inputs line by line from a file.
{Because these operations occur on a shared directory when running on a high-performance computing cluster, file-visibility latency and potential race conditions may arise when multiple processes attempt to access the same files concurrently.
To mitigate these issues, \libyt\ designates the MPI root process as the governing process for all file-based operations; it is the only process permitted to read from and write to the shared directory.
The root process broadcasts commands and gathers results from all other ranks, while the remaining processes do not access the file system and instead communicate exclusively with the root process.
This design avoids race conditions and ensures consistent behavior across nodes.
}
This allows us to reload and update the inline Python function or even run a new Python script. 
Users create specific files to send instructions to \libyt, like when to start reading and executing the file and when to exit the file-based prompt.
The execution results will be output to a file.
See Section \ref{subsec:activating-python-prompt} for demonstration.

\subsection{Jupyter Notebook User Interface} \label{subsec:support-jupyter-notebook}
\libyt\ provides a Python prompt (\sref{subsec:interactive-mode}) for user interaction, but it lacks a user-friendly interface and code syntax highlighting. 
To address these limitations, \libyt\ combines the Project Jupyter \citep[][\url{https://jupyter.org/}]{jupyter} and provides support for Jupyter Notebook and JupyterLab as user interfaces for in situ analysis. 
Hereafter, we refer to them as Jupyter Notebook for convenience.

This Jupyter Notebook access point allows users to connect to simulations running on high-performance computing clusters, access their data, execute Python statements and perform analysis, and use code auto-completion within this well-developed Jupyter Notebook frontend.

\sref{subsubsec:jupyter-project} gives an overview of Project Jupyter and describes how \libyt\ fits into its ecosystem.
Then we describe how we build \libyt\ kernel in \sref{subsubsec:libyt-kernel} and the accompanying Jupyter extension in \sref{subsubsec:libyt-provisioner}.
\sref{subsubsec:kernel-provisioner-jupyter} describes how all the components work together.

\subsubsection{Libyt in Project Jupyter Ecosystem} \label{subsubsec:jupyter-project}
Jupyter Notebook provides an interactive environment where users write and run code, view outputs, and organize everything in a single file.
It is a computational notebook widely used in data analysis and visualization in all industries and research fields.
Behind the scenes, Jupyter Notebook uses a client-server architecture: Jupyter Notebook serves as the frontend, sending user requests (e.g., launching Jupyter kernels, code execution, etc.) to the Jupyter Server, which acts as the central communication hub, including passing Jupyter Notebook requests to the Jupyter Client; Jupyter Client launches and manages the kernel, while the Jupyter kernel is the core process that runs the code and responds to the request initiated by Jupyter Notebook.

The essence of an in situ analysis tool is to avoid moving simulation data and analyze it on-site, so we should create a customized kernel (\sref{subsubsec:libyt-kernel}) that has access to both simulation runtime and Python, and we should create a mechanism to manage our customized kernel (\sref{subsubsec:libyt-provisioner}).
\fref{fig:overview} shows how \libyt\ joins Jupyter Notebook in the architecture alongside other components.

We adopt the Jupyter Notebook user interface to lower the barrier for users.
This feature is an add-on to \libyt.
It provides a better user interface without altering the behavior of other \libyt\ features.

\subsubsection{Building Libyt Kernel} \label{subsubsec:libyt-kernel}
There are many ways to create a Jupyter kernel, such as reusing an existing kernel IPython by overriding the execute request implementation.
For \libyt, it is preferable to implement a kernel with direct access to both the simulation and Python runtimes.
We create \libyt\ kernel at C/C++ runtime using Xeus\footnote{\url{https://xeus.readthedocs.io/}}.
Xeus is a C++ implementation of the Jupyter kernel protocol, 
providing a superclass for kernel developers to inherit and implement their own.
It has a built-in server Xeus-zmq\footnote{\url{https://github.com/jupyter-xeus/xeus-zmq}}, which is the middleware responsible for receiving and sending messages to the Jupyter Client built upon ZeroMQ\footnote{\url{https://zeromq.org/}}.
ZeroMQ is a lightweight messaging library that enables fast asynchronous communication between distributed processes.
We can specify the settings (e.g., ports, IP address, transport method, etc.) or let Xeus-zmq decide automatically.
In either case, a designated file\footnote{\texttt{libyt\_kernel\_connection.json}} specifies the connection information of the \libyt\ kernel.
It may be generated automatically by \libyt\ or supplied manually by the user.

\libyt\ API creates and launches one \libyt\ kernel at the simulation MPI root process, exposing a portal to communicate with Jupyter Client using ZeroMQ. 
The other simulation processes are blocked and wait for the broadcasts from the MPI root process, which serves as the simulation's headquarters.
Thus, we focus on handling requests from Jupyter Client: executing code requests, replying to code-completion requests, handling shutdown requests, and getting kernel information requests.

The workflow of executing code requests follows the same REPL workflow described in \sref{subsubsec:python-prompt-and-libyt-commands}.
Jupyter Client passes the input code via ZeroMQ to the MPI root process, which may contain multiple statements.
After validating the code, the MPI root process globally broadcasts it, and all processes execute it synchronously.
Finally, \libyt\ kernel gathers the output from every MPI process, reformats it to match the Xeus API, and returns the messages to the Jupyter Client. 
Due to the way code is executed and the subsequent output-gathering mechanism, the \libyt\ kernel cannot handle interruptions.
This limitation is discussed in \sref{subsec:limitations}.

For supporting code-completion requests, \libyt\ kernel uses \texttt{jedi}\footnote{\url{https://jedi.readthedocs.io/}}.
\texttt{jedi} is a Python package for Python static analysis with the main feature of auto-completion, and it is widely used in {integrated development environments (IDEs)} and IPython.
\libyt\ kernel delegates the code completion tasks to \texttt{jedi} and passes relevant information (e.g., the code, cursor position, and execution namespace for inspection). 
Then, it returns the results.

When \libyt\ kernel receives a shutdown request, it notifies the other MPI processes, after which all processes terminate.

\libyt\ kernel provides basic information, such as the kernel name and version, the code language and its version, the helper link, etc.
The kernel information setting is directly linked to the Jupyter Notebook frontend, enabling features such as syntax highlighting and the display of kernel documentation.

\subsubsection{Building Libyt Jupyter Extension} \label{subsubsec:libyt-provisioner}

We create \texttt{jupyter\_libyt}\footnote{\url{https://github.com/yt-project/jupyter_libyt}}, a Jupyter extension for \libyt\ kernel.
It involves two parts: the customized backend kernel provisioner (\libyt\ provisioner) to handle \libyt\ kernel and the configuration of the Jupyter Notebook frontend when connecting to it.

Generally, when we launch Jupyter Notebook, the Jupyter Client initializes a kernel process by delegating kernel creation to the default kernel provisioner, which starts the Jupyter kernel and returns its information.
In our case, this order is reversed: we first launch the \libyt\ kernel and then instruct the Jupyter Client to connect to it.

The \libyt\ provisioner manages the simulation-launched \libyt\ kernel by retrieving its connection details (e.g., binding ports, IP address, network protocols, process identifier, etc.).
It obtains the necessary information from files\footnote{\texttt{libyt\_kernel\_pid.txt} tracks the kernel's status, while \texttt{libyt\_kernel\_connection.json} contains the connection configuration.} within the directory designated by the environment variable\footnote{\texttt{LIBYT\_KERNEL\_INFO\_DIR}}.
Users must set this environment variable to the correct directory before starting Jupyter Notebook.
\libyt\ provisioner will not change the original behavior of Jupyter Notebook and is only active when the Jupyter Notebook chooses \libyt\ kernel.

\libyt\ kernel doesn't support interrupting and restarting, including the ``interrupt'', ``restart'', and ``restart and run all'' commands on the toolbar buttons and the kernel main menu.
They are blocked to prevent misleading messages and unintended actions when the \libyt\ kernel is in use.

To make Jupyter recognize \libyt\ kernel, we register a kernel specification file under Jupyter search path\footnote{Environment variable \texttt{JUPYTER\_PATH}.} during \libyt\ installation.

\subsubsection{Combining Libyt Kernel, Libyt Jupyter Extension, and Project Jupyter} \label{subsubsec:kernel-provisioner-jupyter}
Simulations and Jupyter Notebook are run in separate processes, and they are treated individually.
Simulation uses \libyt\ API to create and launch \libyt\ kernel, which establishes an entry point for Jupyter Client to connect to. 
The API also controls whether the connection information is set automatically or manually.
The latter is crucial when submitting jobs in high-performance computing clusters.
After setting the environment variable to the directory where the \libyt\ kernel writes the connection information file (\sref{subsubsec:libyt-provisioner}), the user launches Jupyter Notebook and selects ``Libyt'' from the context menu. 
Jupyter Notebook connects to \libyt\ kernel using Jupyter Client, and it, in turn, delegates the managing kernel jobs to \libyt\ provisioner.


Generally, all functionalities in the Jupyter Notebook frontend are supported and remain the same, except for those related to the kernel.
We can type Python code and press ``Shift + Enter'' to run it, and it will print the results from each MPI process.
We can also use \libyt\ magic commands (\sref{subsubsec:python-prompt-and-libyt-commands}) to toggle the Python function on or off, view function definition, retrieve error messages, etc.
We can use the ``Tab'' key to do auto-completion.
We can save and export the notebook to other formats.
Note that the buttons for controlling kernels are different, as Jupyter Client no longer manages the kernel process itself; instead, the simulation process handles it separately.
The ``Shutdown'' command closes the \libyt\ kernel and returns the processes to simulation.
The simulation will continue its original computing tasks. 
The ``Restart'' and ``Interrupt'' commands are not supported.
{Furthermore, disconnecting from the notebook terminates the \libyt\ kernel, so reconnecting to the same kernel is not currently supported.
Multiple notebooks can attach to the same \libyt\ kernel and execute code within the same namespace (\sref{subsubsec:importing-and-calling-python}).
The \libyt\ kernel processes the code sequentially in the order received.
However, a shutdown initiated by any user terminates the kernel for all users.
}

This feature functions like a standard Jupyter Notebook, but with access to ongoing simulation data, and it does not change the underlying behavior of \libyt.
It supports both serial and parallel computing under MPI, and is also applicable to high-performance computing.

\subsection{Connecting Libyt to yt} \label{subsec:connecting-libyt-to-yt}
This paper focuses on using \yt\ as the analysis package in the in situ Python workflow.
\yt\ provides functionalities such as data object selection methods based on spatial location, filtering methods based on values, calculation of physical quantities for an object, visualization, and plotting of correlations between data, among others.
The field data can be a simulation's native field or a derived field (referred to as ``\yt\ derived field'') that is combined from a set of native fields defined in \yt.
Simulations can also define their unique fields.
\yt\ supports computing most of these operations in parallel using \texttt{mpi4py} \citep[][\url{https://mpi4py.readthedocs.io/}]{mpi4py}.
\texttt{mpi4py} is a Python package that provides Python bindings to the MPI standard, enabling Python applications to run in parallel within the MPI task space.
The job decomposition strategies used in \yt\ are (i) spatial decomposition, which splits the jobs based on the target domain; 
(ii) grid decomposition, {which partitions data objects to be read and processed into smaller chunks, enabling parallel file access and improving data loading performance from disk}; 
and (iii) embarrassingly parallelism, which applies an identical workflow to a group of independent datasets.

Each simulation has its own \yt\ frontend, allowing \yt\ to interpret and process datasets from different simulations.
A \yt\ frontend abstracts a simulation's data structures, metadata, and data fields.
It reads \yt\ parameters, defines simulation-specific parameters, defines simulation native and derived fields in addition to global \yt\ derived fields, loads data structures like AMR grids (\sref{subsec:amr}), and provides methods for querying and getting simulation data snapshots.
Hiding simulation-specific metadata and unifying the frontend for loading data make the analysis process interchangeable between different simulations, making the operation agnostic to the dataset.

We create \texttt{yt\_libyt}\footnote{\url{https://github.com/data-exp-lab/yt_libyt}}, a \yt\ frontend for \libyt.
It is implemented as a standalone package so users can adopt different versions of \yt\ according to their needs. 
{It loads simulation information and provides methods for \yt\ to read and request field data from \libyt\ Python Module (see \sref{subsec:embedding-python-in-c++-application}).}
This is very similar to how other simulation codes implement \yt\ frontends.
{However, in this case, \libyt\ supplies data to \yt\ directly from in-memory datasets distributed across multiple computing nodes, rather than reading from files on disk.}

The way \libyt\ stores the AMR grid structure is aligned with the definition in \yt\ (see \sref{subsec:amr}).
Currently, \libyt\ only supports this data structure.
When loading the structural hierarchy of the grids and setting \yt\ parameters, \texttt{yt\_libyt} reads the \texttt{hierarchy} and \texttt{param\_yt} dictionaries from the \libyt\ Python Module and does not make copies unless special objects need to be created.
For setting code-specific parameters, \texttt{yt\_libyt} adds additional attributes using the key-value pairs in \libyt\ Python Module \texttt{param\_user} dictionary.
This step is crucial when borrowing a simulation frontend's field and particle information, as a simulation frontend may have its own definitions.

\libyt\ can access the field information of \yt\ derived fields.
Moreover, when applying \libyt\ to a simulation code already with a \yt\ frontend for post-processing, it will inherit the field information (e.g., units, alias names) from its \yt\ frontend and append new field definitions added through \libyt\ API if the fields are not defined.
See demonstration in \sref{subsec:adding-derived-fields}.
This also applies to particle information.
Details of how to access \yt\ derived fields, simulation fields defined in its frontend, and \libyt\ fields are described in \sref{subsec:performing-in-situ-analysis}.

Based on the location of the data sample within a unit cell, it can be cell-centered, meaning the center of the unit cell, or face-centered, meaning the center of the cube's surface.
Typical examples of cell-centered fields are gas mass density and gravitational potential in a gravitational hydrodynamic code. 
An example of a face-centered field is the magnetic field defined in the constrained transport algorithm used in magnetohydrodynamics (MHD) \citep[e.g.,][]{Athena}.
\texttt{yt\_libyt} reads cell-centered and face-centered data from the \libyt\ Python Module's \texttt{grid\_data} dictionary; it also reads particle data from \texttt{particle\_data} dictionary.
\yt\ only operates on cell-centered data; thus, we need to ensure face-centered fields are properly converted with minimal memory overhead.
The procedure for obtaining cell-centered or face-centered data, with or without ghost cells, is as follows:
\begin{enumerate}
   \item Read the entire data array from the Python dictionary object \texttt{grid\_data}.
   \item Get the target subarray through NumPy indexing, excluding ghost cells when present.
   \item For face-centered data, convert it to cell-centered data by taking the average over neighboring data along the direction normal to the target face.
\end{enumerate}

For derived fields and particles that require further preparation by the simulation, \texttt{yt\_libyt} invokes the \libyt\ Python Module's C-extension methods.
Details on user-defined functions binding for AMR simulations are given in \sref{subsec:amr}.



Unlike intrinsic fields, computing derived fields requires allocating additional memory, which can be memory-intensive when preparing all grids simultaneously. 
To alleviate this potential issue, \libyt\ only prepares the data requested by \yt.
For example, operations like slice plots only require a small portion of the grid data where the slice intersects.
Furthermore, \libyt\ leverages the data chunking mechanism in \yt.
Before actually reading data from disk or memory, \yt\ identifies the data it needs, groups them into chunks, and then reads and computes one chunk at a time. 
As a result, \libyt\ does not have to prepare all the required data for a given analysis at once.

When getting data not owned by the simulation MPI process, \texttt{yt\_libyt} delegates the data redistribution process to \libyt\ Python Module C-extension methods \texttt{get\_field\_remote} and \texttt{get\_particle\_remote}.
These steps involve collective MPI calls, requiring every MPI process to participate.
The details are described in \sref{subsec:parallelism}.
{Because the need for redistribution is determined only after synchronization, data I/O from the \libyt\ Python module in \texttt{yt\_libyt} is implemented as an MPI collective operation, requiring participation from all processes.}

\texttt{yt\_libyt} hides data mapping details between \yt\ and \libyt\ Python Module, making the in situ script nearly identical to the post-processing script.
We can reuse the post-processing script and convert it into an in situ analysis script with two lines of change, significantly improving its reusability and flexibility.
We demonstrate this in \sref{subsec:performing-in-situ-analysis}.
\tref{tab:supported-yt-functions} lists the key \yt\ functions supported by \libyt. 
Generally speaking, \libyt\ supports all functions of \yt\ unless there is {an asymmetric data-reading phase} in different MPI processes.
We elaborate on this minor limitation in \sref{subsec:limitations}.

\begin{deluxetable}{ll}[htb]
\tablewidth{\columnwidth}
\tablecaption{Supported key \yt\ functions\label{tab:supported-yt-functions}}
\tablehead{
Functions & Limitations
}
\startdata
   Derived quantity & \\
   Projection plot & \\
   Slice plot & \\
   Extract uniform grids & \\
   1D/2D profile & \\
   Line plot & \\
   Annotations & Some functions require all MPI processes \\
               & to call \texttt{save} when saving a figure. \\
   Volume rendering & Number of MPI processes must be even. \\
   Particle data & \\
   Data objects & \\
\enddata
\tablecomments{Generally speaking, \libyt\ supports every \yt\ function
that involves all MPI processes in {data-reading phase}.
See \sref{subsec:limitations} for details.}
\end{deluxetable}

\subsection{Adaptive Mesh Refinement Data Structure} \label{subsec:amr}

{\libyt\ currently supports only patch-based AMR grid data structures; therefore, 
in the following sections, the term ``AMR grid'' refers specifically to a patch-based AMR grid.
}

Section \ref{subsubsec:amr-grid} describes how we define AMR grid data structures.
Section \ref{subsubsec:field-information} and Section \ref{subsubsec:particle-information} illustrate how we deal with fields and particles, respectively.
The mechanism for binding simulation data to Python and enabling back-communication from Python to simulations is described in \sref{subsec:embedding-python-in-c++-application}.

These registered data and functions are mapped to \yt\ through \texttt{yt\_libyt} (\libyt\ frontend for \yt\ in \sref{subsec:connecting-libyt-to-yt}), making simulations accessible in Python and thus enabling in situ analysis.

\subsubsection{Adaptive Mesh Refinement Grid} \label{subsubsec:amr-grid}

In {patch-based} AMR simulations, the computational domain consists of grids with different sizes and resolutions. 
An AMR grid consists of a globally unique ID, its parent grid ID, grid dimensions, left and right edges of the grid, and the refinement level.
The collection of information from every grid forms the hierarchy.
\yt\ requires the whole simulation AMR grid information, but not all AMR simulation codes redundantly store the complete grid hierarchy in each MPI process. 
For example, \gamer\ \citep{gamer-2} only stores the information of the local grid hierarchy associated with the current MPI process.
In this case, \libyt\ helps collect hierarchy information scattered in different MPI processes, so that each process only needs to prepare its local hierarchy.
\libyt\ binds the hierarchy to \libyt\ Python Module \texttt{hierarchy} dictionary.
Each element in the hierarchy is stored in a continuous array, so that \texttt{yt\_libyt} can access them directly without additional translation to \yt\ format.
Since they are universal in all AMR simulations, it does not require the simulation to be written in a specific format to use \libyt.

Depending on the simulation settings and the governing physics, different sets of fields and particles can exist, and particles can be optional.
We can categorize data into two types:
\begin{itemize}
    \item \emph{intrinsic} data that is actual runtime data operated and managed by simulations and belongs to a globally unique grid ID, and
    \item \emph{derived} data that requires further manipulation since the data doesn't exist yet or the format isn't suitable for in situ analysis.
\end{itemize}
Whether intrinsic or derived, they can be retrieved using this grid ID and label names (e.g., field name, particle type, etc.), either by reading \libyt\ Python Module dictionary or by using C-extension methods.

\libyt\ API only binds intrinsic field data and particle data owned by this MPI process.
See \sref{subsec:setting-local-grids-information} for a demonstration.
The following sections deal with the field and particle information.

\subsubsection{Field and Field Derived Function} \label{subsubsec:field-information}

Intrinsic field data is stored in a contiguous array.
Some simulation codes permanently allocate ghost cells to each grid (e.g., \enzo), while others do not (e.g., \gamer).
\libyt\ defines field information such as field name, field type (e.g., cell-centered, face-centered, or derived field), ghost cells, whether the field data is contiguous in the x-axis, field unit, derived field function, etc.
They are essential for \libyt\ to bind underlying field data to \libyt\ Python Module \texttt{grid\_data} dictionary and for \texttt{yt\_libyt} to map them to \yt.
All field data are stored in the nested dictionary with grid IDs and field names as keys.
After completing the in situ analysis, these data will not be modified or deleted so that the simulation can continue without interruption.


Derived fields refer to new field data derived from intrinsic fields.
In other words, they are not stored in memory during simulation runtime and require additional manipulation on intrinsic fields to generate the data.
For example, the original face-centered magnetic field is intrinsic, whereas the corresponding cell-centered magnetic field is derived.
Another example is the conversion between various cell-centered hydrodynamic and thermodynamic variables. 
This conversion can be either straightforward, such as computing velocity from momentum and mass density, or complicated, such as computing temperature from gas mass density, internal energy, and electron fraction for a nuclear equation of state \citep[e.g.,][]{NuclearEoS}, which requires a table lookup.
\libyt\ provides hooks for the user-defined C function to compute a general derived field.
This C function should be able to generate and write data to the provided storage using only the grid ID and field name.
The function is part of the derived field generating routine in \libyt\ Python Module \texttt{derived\_func} C-extension method, done in the following steps:
\begin{enumerate}
   \item Python C-extension method \texttt{derived\_func} allocates data arrays to store derived fields.
   \item Invoke user-defined C function to write the derived field data in the data arrays.
   \item Use NumPy API to wrap around this data array and return to Python.
\end{enumerate}
This C-extension method is called only when the specified derived data is needed, thus enabling back-communication from Python to simulations.
The generated data gets freed once Python no longer needs it.
See \sref{subsec:adding-derived-fields} for a demonstration.

\subsubsection{Particle and Get Particle Function} \label{subsubsec:particle-information}

One key difference between particle and field data is that particle data may not always be contiguous in memory.
Additionally, there can be multiple particle types, each with distinct attributes. 
For example, ``star'' particles can have metallicity, ``black hole'' particles can have spins, and ``dark matter'' particles can have a softening length.
To support this diversity, we define particle metadata such as particle type, total attribute count, attribute names, and the function used to retrieve particle data.
They are required for \libyt\ bind particle data and then map to \yt.

We load particle data based on how it is stored in memory.
For intrinsic particle data stored in a contiguous array, we organize them under \libyt\ Python Module \texttt{particle\_data} dictionary with grid IDs, particle type, and attribute as keys.
For intrinsic particle attributes that are stored sparsely, \libyt\ first allocates a data array and then gathers scattered particle data through a user-provided C function.
This function can also generate derived particle attributes, provided it can generate data based on grid ID, particle type, and attribute.
It is similar to the procedure of computing derived fields as described in the previous subsection.
\libyt\ Python Module C-extension method \texttt{get\_particle} prepares particle data through the following steps:
\begin{enumerate}
   \item Allocate storage for particle data array.
   \item Invoke user-defined C function to write particle data to this array.
   \item Use NumPy API to wrap around this data array and return to Python.
\end{enumerate}
To reduce memory overhead, the C-extension method is called only when the particle data is needed, and the generated data is freed once it is no longer in use.
See \sref{subsec:setting-particle-information} for a demonstration.

\subsection{Parallelism} \label{subsec:parallelism}
\yt\ supports MPI parallelism (see \sref{subsec:connecting-libyt-to-yt}), and it is desirable for \libyt\ to utilize it directly.
There is, however, one critical caveat. 
For post-processing, the grid and particle data stored on hard disks are accessible by all MPI processes, similar to a shared-memory mechanism. 
\yt\ can thus distribute data and associated computation to different MPI processes arbitrarily without requiring additional MPI communication.
How \yt\ distributes these data generally differs from the distribution of in-memory data during simulation runtime.
The former is based on the job decomposition algorithm, which may vary across operations and might not consider data locality or which MPI process holds the data.
In other words, there is no guarantee that the data requested by \yt\ in each MPI process is already located in the same process when conducting in situ analysis.
In a more general case, particularly when using other Python packages that use MPI for parallel computation, we do not want to restrict the in situ Python analysis workflow to parallelism algorithms based on data locality.
Thus, \libyt\ performs additional MPI communication to redistribute grid and particle data in accordance with the job decomposition scheme used by the parallel Python packages.

Many common MPI routines need handshaking between a sender and a receiver.
However, since \libyt\ cannot know in advance what kind of communication pattern a Python script needs, it is difficult to schedule point-to-point communications that fit any algorithms and any number of MPI processes. 
To solve this problem, \libyt\ uses one-sided communication in MPI, also known as Remote Memory Access (RMA), which allows one to avoid explicitly specifying senders and receivers.
By creating an RMA ``epoch'' that all MPI processes enter, each process can access data from other processes without explicitly requiring their participation.
Note that the RMA synchronization routine \texttt{MPI\_Win\_fence}, which starts and ends an epoch, is a collective operation requiring all processes to participate. 
If, however, only a subset of MPI processes is involved in this stage, the RMA operation will hang and fail. 
{We measure the weak and strong scaling of \libyt\ data redistribution process in \sref{subsubsec:weak-strong-scaling-libyt} to assess the impact of collective operations on scalability.}
We will discuss this limitation further in \sref{subsec:limitations}.

Different kinds of unit data may be involved in this data redistribution. 
Unit data refers to the smallest data pieces that \libyt\ operates on.
For example, an AMR grid has a metadata grid ID, data type, data dimensions, data pointers, and a contiguous indicator in the x-axis; and a particle array has a grid ID, data type, data pointer, and length.
\libyt\ redistributes the same kind of unit data at a time in an epoch.
\libyt\ uses the Template Method design pattern to construct a process that applies to all sorts of unit data.
The unit data in consideration should provide the custom MPI data type to gather its metadata, the data buffer size in bytes, and the buffer length, which is the number of elements in the buffer. 
The whole process of this one-sided communication can be summarized as follows:
\begin{enumerate}
   \item Create an MPI window that allows memory to be dynamically exposed and unexposed.
   \item Attach prepared data buffers to the MPI window. Then, get the address of the attached buffer.
   \item Gather the RMA information, such as address and MPI process number; gather the metadata of each unit data.
   \item Open the RMA epoch to start the data redistribution process.
   \item Fetch data from remote processes.
   \item Close the RMA epoch to end the data transfer process.
   \item Detach the data buffer and free data prepared specifically for other processes.
   \item Close the MPI window.
\end{enumerate}

\libyt\ Python Module (\sref{subsec:embedding-python-in-c++-application}) C-extension methods \texttt{get\_field\_remote} and \texttt{get\_particle\_remote} use this data redistribution routine for field and particle data, respectively.
These C-extension methods are collective operations; every MPI process must enter this stage.
Before calling the C-extension methods, Python first gathers the grid IDs requested by Python in each MPI process, so that all processes know which data to prepare for others and which MPI process to request data from.
The data array fetched from other MPI processes is stored in a dictionary and then returned to the caller.

In large simulations, the data communication size may exceed the maximum number of data elements allowed to be transferred at a time\footnote{The upper limit is $2^{31}$ since the MPI parameter \texttt{count} uses the 32-bit integer data type.}.
To overcome this limitation, \libyt\ splits the data transfer of large arrays into multiple MPI calls automatically to ensure that each transfer does not exceed the upper limit.

\section{Demonstration} \label{sec:demonstration}

\begin{figure*}[ht]
\includegraphics[width=\textwidth]{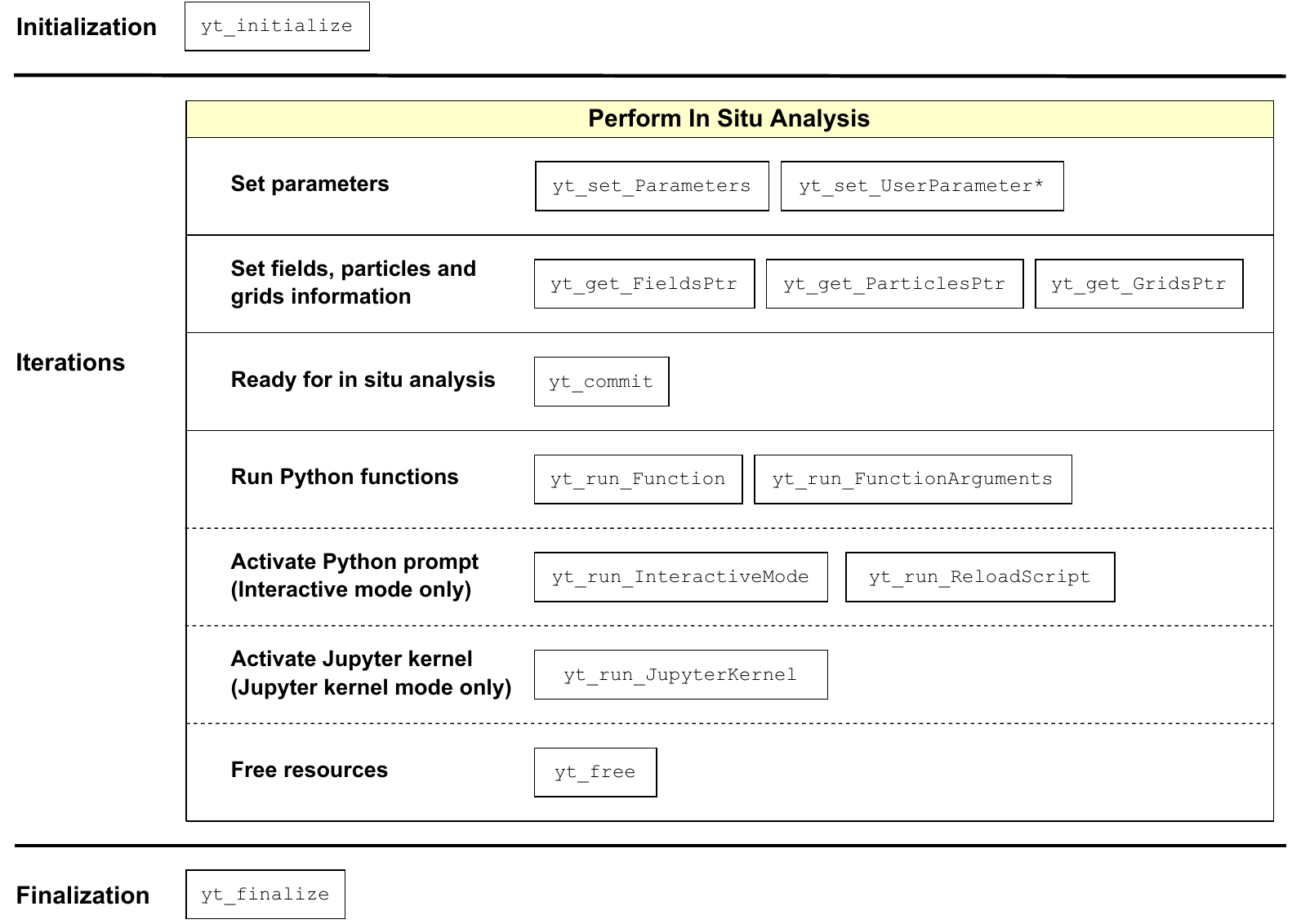}
\caption{
The procedure of performing in situ analysis with \libyt\ and the API related to each step. 
These steps must be followed in every round of the analysis.
Note that the simulation parameters (e.g., current time), grid distribution (e.g., with AMR), and particles associated with each grid can change during simulation runtime; therefore, they are reset at the end of each simulation iteration. 
Some entry points (e.g., Python prompt and Jupyter kernel) may be unavailable based on the compilation options.
\label{fig:procedure}}
\end{figure*}

In this section, we provide code snippets in C++ to demonstrate how to implement \libyt\ into a simulation code and how to invoke Python routines for in situ analysis.
\fref{fig:procedure} shows the procedure of using \libyt\ for in situ analysis during simulation runtime.
\tref{tab:libyt-api} lists all the corresponding functions provided by the \libyt\ API.

\begin{deluxetable*}{lll}
\tablecaption{\libyt\ API\label{tab:libyt-api}}
\tablehead{
\colhead{API} & \colhead{Usage} & \colhead{Section}
}
\startdata
    \texttt{yt\_initialize} & Initialize \libyt. & Section \ref{subsec:initialization} \\
    \texttt{yt\_set\_Parameters} & Set \yt\ parameters. & Section \ref{subsec:setting-parameters} \\
    \texttt{yt\_set\_UserParameter*} & Set code-specific parameters. & Section \ref{subsec:setting-parameters} \\
    \texttt{yt\_get\_FieldsPtr} & Fill in field information. & Section \ref{subsec:setting-field-information} \\
    \texttt{yt\_get\_ParticlesPtr} &  Fill in particle information. & Section \ref{subsec:setting-particle-information} \\
    \texttt{yt\_get\_GridsPtr} & Fill in grid information and data. & Section \ref{subsec:setting-local-grids-information} \\
    \texttt{yt\_commit} & Commit settings before performing in situ analysis. & Section \ref{subsec:committing-settings} \\
    \texttt{yt\_getGridInfo\_*} & Look up hierarchy and data functions. & Section \ref{subsec:look-up-function} \\
    \texttt{yt\_run\_Function} & Call a Python function without arguments. & Section \ref{subsec:performing-in-situ-analysis} \\
    \texttt{yt\_run\_FunctionArguments} & Call a Python function with arguments. & Section \ref{subsec:performing-in-situ-analysis} \\
    \texttt{yt\_run\_InteractiveMode} & Activate Python prompt. (Interactive mode only.) & Section \ref{subsec:activating-python-prompt}  \\
    \texttt{yt\_run\_ReloadScript} & Activate file-based Python prompt and reload script. (Interactive mode only.) & Section \ref{subsec:activating-python-prompt}  \\
    \texttt{yt\_run\_JupyterKernel} & Activate libyt kernel. (Jupyter kernel mode only.) & Section \ref{subsec:activate-libyt-kernel} \\
    \texttt{yt\_free} & Free the resources allocated by \libyt. & Section \ref{subsec:freeing-resources} \\
    \texttt{yt\_finalize} & Finalize \libyt. & Section \ref{subsec:finalization}
\enddata
\end{deluxetable*}

\subsection{Initialization} \label{subsec:initialization}

At the initialization stage, we set the general configuration of \libyt\ such as the logging level and debugging mode, and import a user-provided Python script for in situ analysis later.
The debugging mode checks, for example, whether field information, particle information, and the AMR grid hierarchy are set correctly.
We also initialize the Python interpreter and \libyt\ Python Module at this stage,
which only needs to be done once.

The following example imports the Python script \texttt{inline\_script.py}, disables both logging and debugging, and initializes \libyt\ by calling the API function \texttt{yt\_initialize}.
\\
\\
{\small
\begin{tabular*}{\columnwidth}{l}
    \toprule
    \texttt{yt\_param\_libyt param\_libyt;} \\
    \texttt{param\_libyt.script = "inline\_script";} \\
    \texttt{param\_libyt.verbose = YT\_VERBOSE\_OFF;} \\
    \texttt{param\_libyt.check\_data = false;} \\
    \texttt{yt\_initialize( argc, argv, \&param\_libyt );} \\
    \hline
\end{tabular*}
}

\subsection{Setting Parameters} \label{subsec:setting-parameters}

In the following example, we store all \yt-specific parameters in \texttt{param\_yt} and pass them to \yt\ by calling the API function \texttt{yt\_set\_Parameters}.
These parameters include, for instance, \texttt{frontend} (name of the corresponding \yt\ frontend), \texttt{current\_time} (current simulation time), \texttt{time\_unit} (time unit in seconds), \texttt{dimensionality} (domain dimensionality), \texttt{num\_fields} (number of fields), \texttt{num\_par\_types} (number of particle types), \texttt{num\_attr} (number of particle attributes associated with each type), \texttt{num\_grids\_local} (number of local grids on this MPI process), and \texttt{index\_offset} (grid index offset).
Note that we use \gamer\ as an example, which already has a corresponding \yt\ frontend. 
Hence, we can directly borrow the field information defined in this frontend without the need to redefine it here (see \sref{subsec:connecting-libyt-to-yt}).
\\
\\
{\small
\begin{tabular*}{\columnwidth}{l}
    \toprule
    \texttt{yt\_param\_yt param\_yt;} \\
    \texttt{param\_yt.frontend = "gamer";} \\
    \texttt{param\_yt.current\_time = 0.0;} \\
    \texttt{param\_yt.time\_unit = 3.1557e13;} \\
    \texttt{param\_yt.dimensionality = 3;} \\
    \texttt{param\_yt.num\_fields = 5;} \\
    \texttt{param\_yt.num\_grids\_local = 100;} \\
    \texttt{param\_yt.index\_offset = 0;} \\
    \texttt{/* Set particle types. */} \\
    \texttt{param\_yt.num\_par\_types = 1;} \\
    \texttt{yt\_par\_type par\_type\_list[num\_par\_types];} \\
    \texttt{par\_type\_list[0].par\_type = "io";} \\
    \texttt{par\_type\_list[0].num\_attr = 4;} \\
    \texttt{param\_yt.par\_type\_list = par\_type\_list;} \\
    \texttt{yt\_set\_Parameters( \&param\_yt );} \\
    \hline
\end{tabular*}
}
\\

Simulation codes can also define their own parameters and pass them to \yt\ by calling the API function \texttt{yt\_set\_UserParameter*}, where the asterisk \texttt{*} shall be replaced by \texttt{Int}, \texttt{Long}, \texttt{LongLong}, \texttt{Uint}, \texttt{Ulong}, \texttt{Float}, \texttt{Double}, or \texttt{String} depending on the input data type.
The following example adds a single integer parameter \texttt{mhd} to the \gamer\ frontend, indicating the disabling of MHD.
\\
\\
{\small
\begin{tabular*}{\columnwidth}{l}
    \toprule
    \texttt{const int MHD = 0;} \\
    \texttt{y\_set\_UserParameterInt( "mhd", 1, \&MHD ); } \\
    \hline
\end{tabular*}
}

\subsection{Setting Field Information} \label{subsec:setting-field-information}

To set the field information, such as field name, unit, and data format, the simulation code first calls the API function \texttt{yt\_get\_FieldsPtr} to obtain the pointer to \texttt{yt\_field} array and then fills in the information for each field.
The following example assumes that the first field is gas mass density named \texttt{Dens}, which is cell-centered, in the single-precision floating-point data format (\texttt{YT\_FLOAT}), and contiguous in memory along the x-axis (\texttt{contiguous\_in\_x = true}).
Note that \libyt\ also supports continuous data along the z-axis by setting \texttt{contiguous\_in\_x = false}. 
We omit specifying the field unit because the field name ``Dens'' is already defined in \gamer's \yt\ frontend.
\libyt\ can therefore utilize the field information directly.
By default, \libyt\ assumes no ghost cell.
\\
\\
{\small
\begin{tabular*}{\columnwidth}{l}
    \toprule
    \texttt{yt\_field *fields;} \\
    \texttt{yt\_get\_FieldsPtr( \&fields );} \\
    \texttt{fields[0].field\_name = "Dens";} \\
    \texttt{fields[0].field\_type = "cell-centered";} \\
    \texttt{fields[0].field\_dtype = YT\_FLOAT;} \\
    \texttt{fields[0].contiguous\_in\_x = true;} \\
    \hline
\end{tabular*}
}

\subsection{Adding Derived Fields} \label{subsec:adding-derived-fields}

We have introduced derived fields in \sref{subsubsec:field-information}.
The procedure of adding a derived field is similar to registering an intrinsic field (\sref{subsec:setting-field-information}) except that a simulation code must provide a data-generating function to construct each derived field.
The following example creates a derived field ``Half\_Dens'', which divides the original density field by $2$, by setting \texttt{field\_type} to \texttt{derived\_func} and providing a data-generating C function \texttt{half\_dens}. 
Because this new field is not defined under \gamer's \yt\ frontend, we provide the field unit by setting \texttt{field\_unit}.
Other parameters are the same as setting the density intrinsic fields.
\\
\\
{\small
\begin{tabular*}{\columnwidth}{l}
    \toprule
    \texttt{void half\_dens(const int len, } \\
    \texttt{const long *gids, const char *field,} \\
    \texttt{yt\_array *output) \{} \\
    \texttt{\quad for (int id=0; id<len; id++) \{} \\
    \texttt{\quad \quad yt\_data dens;} \\
    \texttt{\quad \quad yt\_getGridInfo\_FieldData(gids[id], } \\
    \texttt{\quad \quad \quad "Dens", \&dens);} \\
    \texttt{\quad \quad int dens\_len = output[id].data\_length;} \\
    \texttt{\quad \quad for (int index=0; index<dens\_len;} \\
    \texttt{\quad \quad \quad index++) \{} \\
    \texttt{\quad \quad \quad ((float*)output[id].data\_ptr)[index] = } \\
    \texttt{\quad \quad \quad \quad ((float*)dens.data\_ptr)[index] / 2.0;} \\
    \texttt{\quad \quad \}} \\
    \texttt{\quad \}} \\
    \texttt{\}} \\
    \texttt{fields[0].field\_name = "Half\_Dens";} \\
    \texttt{fields[0].field\_type = "derived\_func";} \\
    \texttt{fields[0].derived\_func = half\_dens;} \\
    \texttt{fields[0].field\_dtype = YT\_FLOAT;} \\
    \texttt{fields[0].field\_unit = "code\_mass / code\_length**3";} \\
    \texttt{fields[0].contiguous\_in\_x = true;} \\
    \hline
\end{tabular*}
}
\\

The function prototype of a data-generating function, such as \texttt{half\_dens}, is also given in the above example.
Here \texttt{gids} lists the target grid indices, \texttt{len} is the length of the list, \texttt{field} is the field name, and {\texttt{yt\_data *output}} is the storage for the output derived field data.
This data-generating function should be able to generate derived field data {in single-precision} that has continuous memory along the x-axis and excludes ghost cells based on grid ID and field name. 
It is part of the routine in \libyt\ Python Module C-extension method and is called during the in situ analysis.
{\texttt{half\_dens} first finds the targeting density field using \texttt{yt\_getGridInfo\_FieldData}.}
\libyt\ provides a set of hierarchy and data lookup functions (\sref{subsec:look-up-function}) to assist the data-generating function.
{Because the ``Dens'' field does not contain ghost cells, the ``Half\_Dens'' output corresponds one-to-one to the field values and is computed by iterating over all grid data points and dividing each value by $2$.}
{Note that \texttt{yt\_array} provides the length of \texttt{data\_ptr}; accordingly, we set the iteration range using this value.
}

\subsection{Setting Particle Information} \label{subsec:setting-particle-information}

Setting particle information is similar to setting field information and adding a derived field.
A simulation code first calls the API function \texttt{yt\_get\_ParticlesPtr} to obtain a pointer to the \texttt{yt\_particle} array and then fills in the information for each particle type. 
For flexibility, different particle types can have various attributes, and each attribute can have its own data format (e.g., single-precision floating-point \texttt{YT\_FLOAT}, double-precision floating-point \texttt{YT\_DOUBLE}, integer \texttt{YT\_INT}, etc).
For each particle type, a simulation code must provide a data-generating C function to return the associated particle attribute data if the data is not stored in contiguous memory or if it contains derived attributes.
Users can also supply a continuous particle array directly, as described in \sref{subsec:setting-local-grids-information}.

The following example assumes only one particle type ``io'' (set in \sref{subsec:setting-parameters}) with four attributes ``ParPosX'', ``ParPosY'', ``ParPosZ'', and ``ParMass'' stored in the single-precision floating-point data format. 
\texttt{par\_io\_get\_par\_attr} is the corresponding data-generating C function for all attributes associated with this particle type, which is passed to \libyt\ via setting it to \texttt{get\_par\_attr}.
Note that the particle position attributes are mandatory for all particle types, which can be specified by the parameters \texttt{coor\_x}, \texttt{coor\_y}, \texttt{coor\_z}.
Since all of the attributes are already defined in \gamer's \yt\ frontend, we omit specifying their units and make \libyt\ use the attribute information from the frontend directly.
\\
\\
{\small
\begin{tabular*}{\columnwidth}{l}
    \toprule
    \texttt{void par\_io\_get\_par\_attr(const int len, } \\
    \texttt{const long *gids, const char *par\_type, } \\
    \texttt{const char *attr, yt\_array *output);} \\
    ... \\
    \texttt{yt\_particle *particles;} \\
    \texttt{yt\_get\_ParticlesPtr( \&particles );} \\
    \texttt{const char *attr\_name[]  = \{} \\
    \texttt{\quad "ParPosX", "ParPosY", "ParPosZ", "ParMass"\};} \\
    \texttt{for(int v=0; v<4; v++)\{} \\
    \texttt{\quad particles[0].attr\_list[v].attr\_name  = attr\_name[v];} \\
    \texttt{\quad particles[0].attr\_list[v].attr\_dtype = YT\_FLOAT;} \\
    \texttt{\}} \\
    \texttt{particles[0].coor\_x = attr\_name[0];} \\
    \texttt{particles[0].coor\_y = attr\_name[1];} \\
    \texttt{particles[0].coor\_z = attr\_name[2];} \\
    \texttt{particles[0].get\_par\_attr = par\_io\_get\_par\_attr;} \\
    \hline
\end{tabular*}
}
\\

The function prototype of a particle attribute data-generating C function, such as \texttt{par\_io\_get\_par\_attr}, is also given in the above example, where \texttt{len}, \texttt{gids}, and {\texttt{output}} are the same as the data-generating function for a derived field (\sref{subsec:adding-derived-fields}).
{The parameters \texttt{par\_type} and \texttt{attr} specify the particle type and the target attribute, respectively.}
This function should collect the attribute data in the targeting grid indices listed in \texttt{gids} and store the attributes in {\texttt{output}}.
Note that we can also create \emph{derived} particle attributes using this data-generating C function.
The function can utilize the \libyt\ API lookup hierarchy and data function (\sref{subsec:look-up-function}) to assist in generating these derived particle attributes.

\subsection{Setting Local Grids Information} \label{subsec:setting-local-grids-information}

As described in \sref{subsubsec:amr-grid}, each MPI process only needs to provide the hierarchy information and data of the local grids associated with this process.
The following example first calls the API function \texttt{yt\_get\_GridsPtr} to obtain the pointer to the \texttt{yt\_grid} array, which stores the grid information for each grid held by the simulation process.
It fills in the grid information with a unique \texttt{id} of 0 at the first element of the \texttt{yt\_grid} array (\texttt{grids[0]}), as a demonstration.
{\libyt\ uses $0$-indexed AMR level, where \texttt{level=0} corresponds to the root grid covering the entire simulation domain at a user-specified coarsest resolution, and higher values represent progressively finer grids.}
The parent grid index \texttt{parent\_id} is set to $-1$ since this grid is at the root level and does not have a parent grid.
The field data pointer \texttt{data} points to the in-memory data array of the first field on this grid.
Note that the field data pointers set here and the field information array set in \sref{subsec:setting-field-information} must share the same indexing, meaning each field occupies the same position in both arrays.
We can also assign a particle attribute data array to this grid.
The attribute data pointers assigned to API should align with the particle information array set in \sref{subsec:setting-particle-information}.
For example, \texttt{particle\_data[0][1].data\_ptr = pos\_y} assigns array pointer \texttt{pos\_y} to the second attribute of the first particle type.
We also set the grid's left edge, right edge, and number of cells along each spatial direction.
\\
\\
{\small
\begin{tabular*}{\columnwidth}{l}
    \toprule
    \texttt{yt\_grid *grids;} \\
    \texttt{yt\_get\_GridsPtr( \&grids );} \\
    \texttt{grids[0].id = 0;} \\
    \texttt{grids[0].parent\_id = -1;} \\
    \texttt{grids[0].level = 0;} \\
    \texttt{grids[0].field\_data[0].data\_ptr = data;} \\
    \texttt{grids[0].particle\_data[0][1].data\_ptr = pos\_y;}\\
    \texttt{for(int d=0; d<3; d++)\{} \\
    \texttt{\quad grids[0].left\_edge[d] = 0.0;} \\
    \texttt{\quad grids[0].right\_edge[d] = 1.0;} \\
    \texttt{\quad grids[0].grid\_dimensions[d] = 8;} \\
    \texttt{\}} \\
    \hline
\end{tabular*}
}

\subsection{Committing Settings} \label{subsec:committing-settings}

After setting the aforementioned prerequisite information, a simulation code must call the API function \texttt{yt\_commit} before performing in situ analysis. 
This function registers all information in the \libyt\ Python Module.
\\
\\
{\small
\begin{tabular*}{\columnwidth}{l}
    \toprule
    \texttt{yt\_commit();} \\
    \hline
\end{tabular*}
}

\subsection{Hierarchy and Data Lookup Functions} \label{subsec:look-up-function}

\libyt\ provides hierarchy and data lookup functions for querying the information.
These lookup functions are only available after committing the information, and they are intended for use in user-defined C functions for derived fields (\sref{subsec:adding-derived-fields}) and particles (\sref{subsec:setting-particle-information}) during in situ analysis.
\libyt\ provides API \texttt{yt\_getGridInfo\_*}, where the asterisk \texttt{*} should be substituted by 
\texttt{Dimensions} (for AMR grid dimensions), \texttt{LeftEdge} (for left edge), \texttt{RightEdge} (for right edge), \texttt{ParentId} (for its parent grid ID), \texttt{Level} (for AMR level), \texttt{ProcNum} (for the MPI rank currently holding this grid), \texttt{ParticleCount} (for the number of particles of a given type on this grid), \texttt{FieldData} (for retrieving the field data on this grid), \texttt{ParticleData} (for retrieving the particle data on this grid).
Note that the retrieved field data and particle data pointers from the API are managed and used by the simulations.

\subsection{Performing In Situ Analysis} \label{subsec:performing-in-situ-analysis}

A simulation code can call all the Python functions defined in the Python script loaded at the initialization stage (i.e., \texttt{"inline\_script"} in \sref{subsec:initialization}).
We use the API \texttt{yt\_run\_FunctionArguments} and \texttt{yt\_run\_Function} to execute Python functions with and without arguments, respectively.
The following example illustrates the usage of both cases, where \texttt{func1()} has no argument and \texttt{func2("str\_arg")} has a single string variable \texttt{"str\_arg"}.
\\
\\
{\small
\begin{tabular*}{\columnwidth}{l}
    \toprule
    \texttt{yt\_run\_Function("func1");} \\
    \texttt{yt\_run\_FunctionArguments("func2", 1, "$\backslash$"str\_arg$\backslash$"");} \\
    \hline
\end{tabular*}
}
\\

A Python script for in situ analysis is \emph{very similar} to that for post-processing, which allows \libyt\ to reuse a post-processing script with minimum modifications and thereby significantly lowers the barrier to conducting in situ analysis.
The only differences are that an inline script must import the \texttt{yt\_libyt} package (the \libyt\ frontend; see \sref{subsec:connecting-libyt-to-yt}) and then load data through \texttt{yt\_libyt.libytDataset()}.

The following example demonstrates a post-processing script that generates a projection plot along the x-axis from the density field of the data snapshot \texttt{Data000} and saves the figure to disk on the root MPI process. 
When using multiple MPI processes, the parallelism feature in \yt\ must be enabled by calling \texttt{yt.enable\_parallelism()}.
\\
\\
{\small
\begin{tabular*}{\columnwidth}{l}
    \toprule
    \texttt{import yt} \\
    \texttt{yt.enable\_parallelism()} \\
    \texttt{def func():} \\
    \texttt{\quad \quad ds = yt.load("Data000")} \\
    \texttt{\quad \quad proj = yt.ProjectionPlot(ds, "x", } \\
    \texttt{\quad \quad \quad \quad \quad \quad \quad \quad \quad \quad \quad \quad \quad \quad "density")} \\
    \texttt{\quad \quad if yt.is\_root():} \\
    \texttt{\quad \quad \quad \quad proj.save()} \\
    \hline
\end{tabular*}
}
\\

In comparison, the corresponding in situ analysis script is provided below, which imports \texttt{yt\_libyt} and replaces \texttt{ds=yt.load("Data000")} in the post-processing script with \texttt{ds=yt\_libyt.libytDataset()}.
\\
\\
{\small
\begin{tabular*}{\columnwidth}{l}
    \toprule
    \texttt{import yt\_libyt} \\
    \texttt{import yt} \\
    \texttt{yt.enable\_parallelism()} \\
    \texttt{def func():} \\
    \texttt{\quad \quad ds = yt\_libyt.libytDataset()} \\
    \texttt{\quad \quad proj = yt.ProjectionPlot(ds, "x", } \\
    \texttt{\quad \quad \quad \quad \quad \quad \quad \quad \quad \quad \quad \quad \quad \quad "density")} \\
    \texttt{\quad \quad if yt.is\_root():} \\
    \texttt{\quad \quad \quad \quad proj.save()} \\
    \hline
\end{tabular*}
}
\\



Depending on whether the field definition should be retrieved from the \libyt\ API (\sref{subsec:setting-field-information}) or from \yt\ (e.g., \yt\ derived fields and simulation derived fields), we use the frontend name (specified in \sref{subsec:setting-parameters}) to distinguish between the two sources.
In the \texttt{SlicePlot} example below, the first line \texttt{("gamer", "Dens")} retrieves the simulation data field set via the \libyt\ API, where \texttt{gamer} indicates the frontend name.
The second line \texttt{("gas", "velocity\_x")} retrieves the field definition in \yt\footnote{We can check the known derived fields supported by \yt\ in the current dataset by typing \texttt{ds.derived\_field\_list}, where \texttt{ds} is the dataset loaded.}.
\\
\\
{\small
\begin{tabular*}{\columnwidth}{l}
    \toprule
    \texttt{yt.SlicePlot(ds, "z", ("gamer", "Dens"))} \\
    \texttt{yt.SlicePlot(ds, "z", ("gas", "velocity\_x"))} \\
    \hline
\end{tabular*}
}

\subsection{Activating Python Prompt} \label{subsec:activating-python-prompt}

A simulation code must call the API function \texttt{yt\_run\_InteractiveMode} to activate the Python prompt. 
This API is only available in interactive mode (see \sref{subsec:interactive-mode}).
The terminal at the MPI root process serves as the entry point for the user interface.
It will print out feedback from the Python interpreter and wait for user input. 
We can update our Python functions, define new variables, and treat it like a standard Python prompt, but with access to simulation data.
All loaded Python statements and scripts persist across calls, carrying their state into subsequent in situ analysis steps.

The Python prompt is activated conditionally by detecting the presence of a flag file.
The following example activates the Python prompt if it finds a file named \texttt{LIBYT\_STOP}.
\\
\\
{\small
\begin{tabular*}{\columnwidth}{l}
    \toprule
    \texttt{yt\_run\_InteractiveMode("LIBYT\_STOP");} \\
    \hline
\end{tabular*}
}

The terminal on the MPI root process prints out the status of each Python function and whether the \libyt\ API will invoke them in the following in situ analysis. 
The following indicates that the \libyt\ API has successfully invoked the Python function \texttt{test\_function} and will be automatically executed by \libyt\ in the next round of in situ analysis.
\\
\\
{\small
\begin{tabular*}{\columnwidth}{l}
    \toprule
    \texttt{=================================================} \\
    \texttt{  Inline Function \quad\quad\quad\quad\quad\quad\quad Status \quad\enspace Run} \\
    \texttt{  * test\_function \quad\quad\quad\quad\quad\quad\quad  success \quad\enspace V   } \\
    \texttt{=================================================} \\
    \texttt{>>> } \\
    \hline
\end{tabular*}
}
\\

Note that because \libyt\ grabs user inputs through the prompt, we need to run an interactive job when using this feature in a high-performance computing cluster (e.g., \texttt{qsub -I} for PBS scheduler).

\libyt\ interactive mode also provides a file-based Python prompt, where the user creates specific files to send instructions to \libyt\ and receives output from those files. 
The file-based Python prompt can be used in a high-performance computing cluster and is not restricted to interactive jobs.
The following example enters the file-based Python prompt if a \texttt{LIBYT\_STOP} file is detected or an error occurs when running inline Python functions (\sref{subsec:performing-in-situ-analysis}).
\libyt\ starts reloading the script \texttt{reload.py} once it detects a \texttt{RELOAD} file.
Outputs will be written to either the \texttt{RELOAD\_SUCCESS} or \texttt{RELOAD\_FAILED} file, depending on whether the whole process completes successfully or fails.
To exit the file-based prompt, the user simply creates a \texttt{RELOAD\_EXIT} file.
\\
\\
{\small
\begin{tabular*}{\columnwidth}{l}
    \toprule
    \texttt{yt\_run\_ReloadScript("LIBYT\_STOP", "RELOAD", } \\
    \texttt{\quad \quad \quad \quad \quad \quad \quad \quad \quad \enspace \texttt{"reload.py");}} \\
    \hline
\end{tabular*}
}

\subsection{Activating libyt Kernel} \label{subsec:activate-libyt-kernel}

We can activate \libyt\ kernel in simulations for Jupyter Notebook to connect in Jupyter kernel mode (see \sref{subsec:support-jupyter-notebook}).
This kernel is capable of handling parallel computation using Python syntax via MPI.
It is similar to using a standard Jupyter Notebook with pre-loaded simulation data.
The API allows us to launch the \libyt\ kernel in a remote process, such as a high-performance computing cluster.
Then, we can launch Jupyter Notebook in no-browser mode on the cluster login node and use port forwarding to redirect Jupyter Notebook to our local desktop.
Finally, we can connect to \libyt\ kernel located in the computing node by selecting it in the Jupyter Notebook frontend.

The following example activates a \libyt\ kernel if a \texttt{LIBYT\_STOP} file is detected.
If the second argument is \texttt{true}, it uses a user-provided connection file to set up \libyt\ kernel.
The connection file name must be \texttt{libyt\_kernel\_connection.json}.
If it is \texttt{false}, it automatically selects ports and resorts to default settings.
Generally, we automatically set the configuration on local machines, and we manually set it for simulations running on remote clusters.
\\
\\
{\small
\begin{tabular*}{\columnwidth}{l}
    \toprule
    \texttt{yt\_run\_JupyterKernel("LIBYT\_STOP", true);} \\
    \hline
\end{tabular*}
}

\subsection{Freeing Resources} \label{subsec:freeing-resources}

After completing the in situ analysis in this iteration of the simulation, the simulation code should reset the internal \libyt\ status and release the resources temporarily allocated by \libyt\ via the API function \texttt{yt\_free}.
\\
\\
{\small
\begin{tabular*}{\columnwidth}{l}
    \toprule
    \texttt{yt\_free();} \\
    \hline
\end{tabular*}
}

\subsection{Finalization} \label{subsec:finalization}

When terminating the whole simulation process, the simulation code can clean up the Python interpreter by calling the API function \texttt{yt\_finalize}, which must be done before calling \texttt{MPI\_Finalize}.
\\
\\
{\small
\begin{tabular*}{\columnwidth}{l}
    \toprule
    \texttt{yt\_finalize();} \\
    \hline
\end{tabular*}
}

\section{Applications} \label{sec:applications}

To demonstrate the importance of in situ analysis and the capability of
\libyt, we run several applications, 
from small simulations that can be done on a local desktop to extensive simulations that need to be done in a high-performance computing cluster.

We have implemented \libyt\ into \gamer\ and applied it to three
astrophysical applications: core-collapse supernovae (CCSN; \sref{subsec:ccsn}),
isolated dwarf galaxies (\sref{subsec:dwarf}), and soliton condensation in fuzzy dark matter (FDM; \sref{subsec:fdm}).
We have also implemented \libyt\ into \enzo\ and demonstrated its applicability across one, two, and three-dimensional problems, including
the classical one-dimensional Sod shock tube test (\sref{enzo:sod-shock-tube}), 
the two-dimensional Kelvin-Helmholtz instability problem (\sref{enzo:kelvinhelmholtz}), 
and the three-dimensional AGORA galaxy simulations (\sref{enzo:agora-galaxy}).

\subsection{Core-Collapse Supernovae} \label{subsec:ccsn}

\begin{figure}[ht]
\includegraphics[width=\columnwidth]{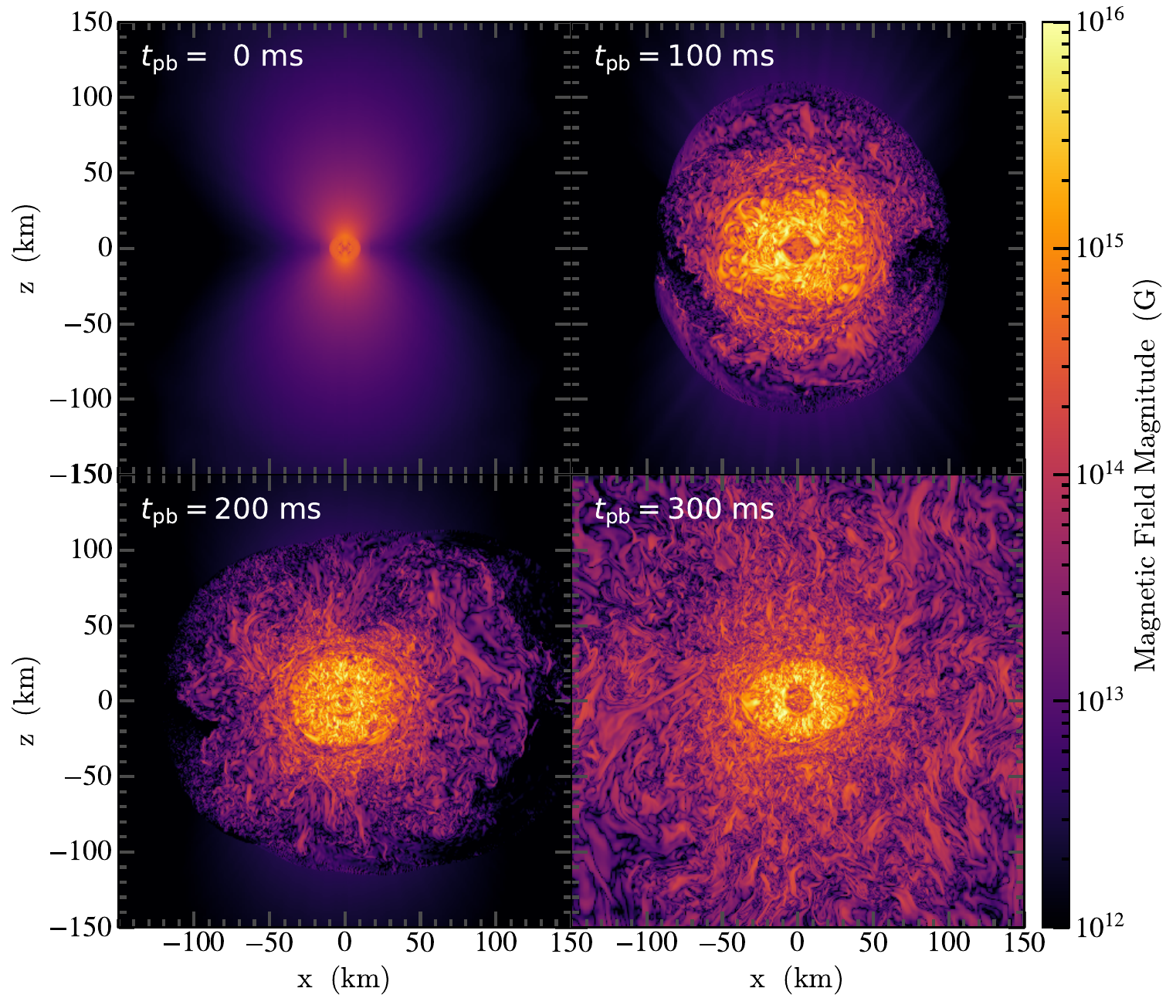}
\caption{
Distribution of magnetic field magnitude at various post-bounce times $t_{\rm pb}$ in a core-collapse supernova simulation performed with \gamer. The visualization data are extracted using the \texttt{save\_as\_dataset} function in \yt{, then} plotted and combined into a 12-second animation. {The animation illustrates the evolution of the magnetic field within the central proto-neutron star and the turbulent post-bounce shock region at high temporal resolution. Shock revival occurs at $t_{\rm pb} \sim 230\ms$. The animation begins at $t_{\rm pb}=-1.16 \ms$ and ends at $t_{\rm pb} = 305.19 \ms$ with a real-time duration of 12 seconds.}
}
\label{fig:ccsn}
\end{figure}

We perform three-dimensional CCSN simulations using \gamer. The code solves the Newtonian magnetohydrodynamics equations using Hancock's Monotonic Upstream-centered Scheme for Conservation Laws \citep[MUSCL;][]{Toro2009}, the HLLD Riemann solver \citep{Miyoshi2005}, and the piecewise parabolic method \citep[PPM;][]{Woodward1984} for data reconstruction. To account for general relativistic effects, we adopt the effective relativistic potential proposed by \citet{2006A&A...445..273M}. Neutrino effects are approximated using a parametrized deleptonization scheme \citep{2005ApJ...633.1042L} during the collapse phase and a leakage scheme \citep{2010CQGra..27k4103O} after core bounce. We employ the finite-temperature, tabulated equation of state with a bulk incompressibility of $K = 220\MeV$ \citep[LS220;][]{1991NuPhA.535..331L}. The simulations follow the dynamical evolution of a $40\Msun$ progenitor from \citet{2007PhR...442..269W}, augmented with an initial rotation profile and a poloidal magnetic field. The spatial resolution is $125\km$ on the root grid with nine refinement levels, achieving a finest resolution of $0.24\km$ to resolve the proto-neutron star ($r \lesssim 30\km$) and post-bounce shock regions. To reduce computational cost, a maximum angular resolution of $0.3^\circ$ is applied to cells located beyond $150\km$ once the proto-neutron star has exploded. Each simulation is evolved to $300\ms$ after core bounce.

Simulation snapshots are output every $2\ms$, with the size increasing from $7.5\GB$ at core bounce to $90\GB$ at the end of the simulation as the shock expands. This output frequency, however, remains insufficient to fully resolve rapidly growing instabilities in three-dimensional, rotating, magnetized progenitors, such as the magnetorotational instability (MRI), the low-$T/|W|$ instability (where $T$ and $W$ denote the rotational and gravitational energies, respectively), and the standing accretion shock instability (SASI). For example, both the MRI and low-$T/|W|$ instabilities evolve on dynamical timescales of $\sim 0.5 \textrm{ -- } 1.0\ms$, shorter than the snapshot interval of $2\ms$. To address this limitation without significantly increasing storage requirements, we use \libyt\ to dump fluid quantities on the $XY$- and $XZ$-planes every $0.2\ms$ via the \texttt{save\_as\_dataset} function in \yt\ for post-processing. \fref{fig:ccsn} shows the magnetic field magnitude on the $XZ$-plane at various post-bounce times $t_{\rm pb}$. Each two-dimensional data file requires only $\sim 50 \textrm{ -- } 200\MB$, offering sufficiently fine temporal resolution to track the dynamical evolution of substructures and instabilities while greatly reducing storage demands. The in situ analysis with \libyt\ costs only $\sim 3 \textrm{ -- } 5\%$ of the total simulation time.
We then further plot and combine these processed datasets into an animation.

\subsection{Isolated Dwarf Galaxies} \label{subsec:dwarf}

\begin{figure}[ht!]
\includegraphics[width=1.0\columnwidth]{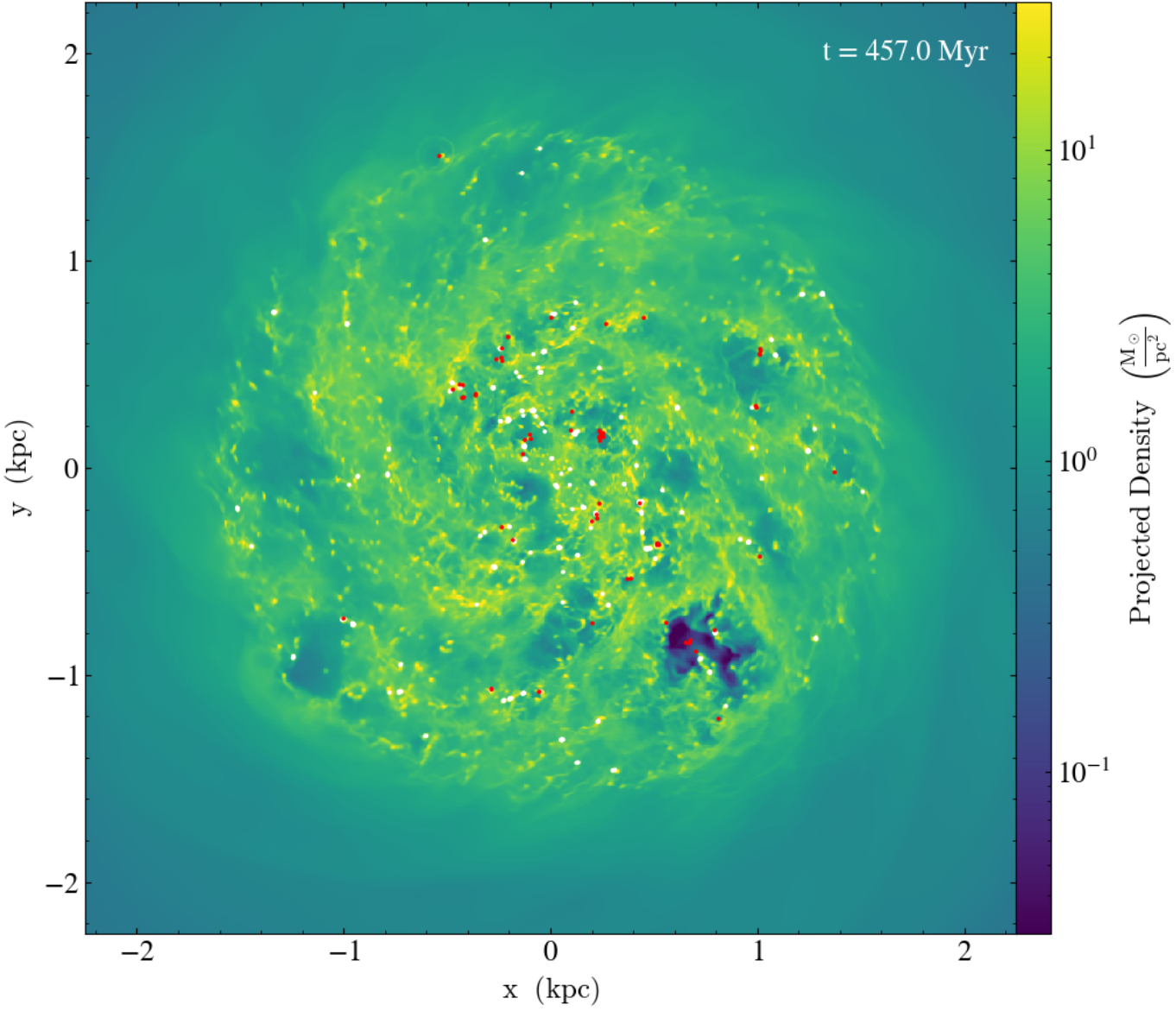}
\caption{
Isolated dwarf galaxy simulation performed with \gamer. {The projected gas density is} annotated with young stars (white dots; age $\le 2.5\Myr$) and recent supernovae (red dots; exploded within the past $2.5\Myr$). {Gas collapses under self-gravity and cools via metal-line radiation. When the Jeans length becomes unresolved, star particles form stochastically and explode as core-collapse supernovae after $10\Myr$, injecting $10^{51}\,\rm erg$ thermal energy and driving blast waves that redistribute gas and regulate star formation.} The dark-blue region near $(x,y) = (0.7\kpc, -0.8\kpc)$ marks a recent strong explosion.
{An animation is available that spans $0\textrm{ -- }750\Myr$ with a real-time duration of 47 seconds.}
}
\label{fig:dwarf-galaxy_slice}
\end{figure}

\begin{figure}[ht!]
\includegraphics[width=1.0\columnwidth]{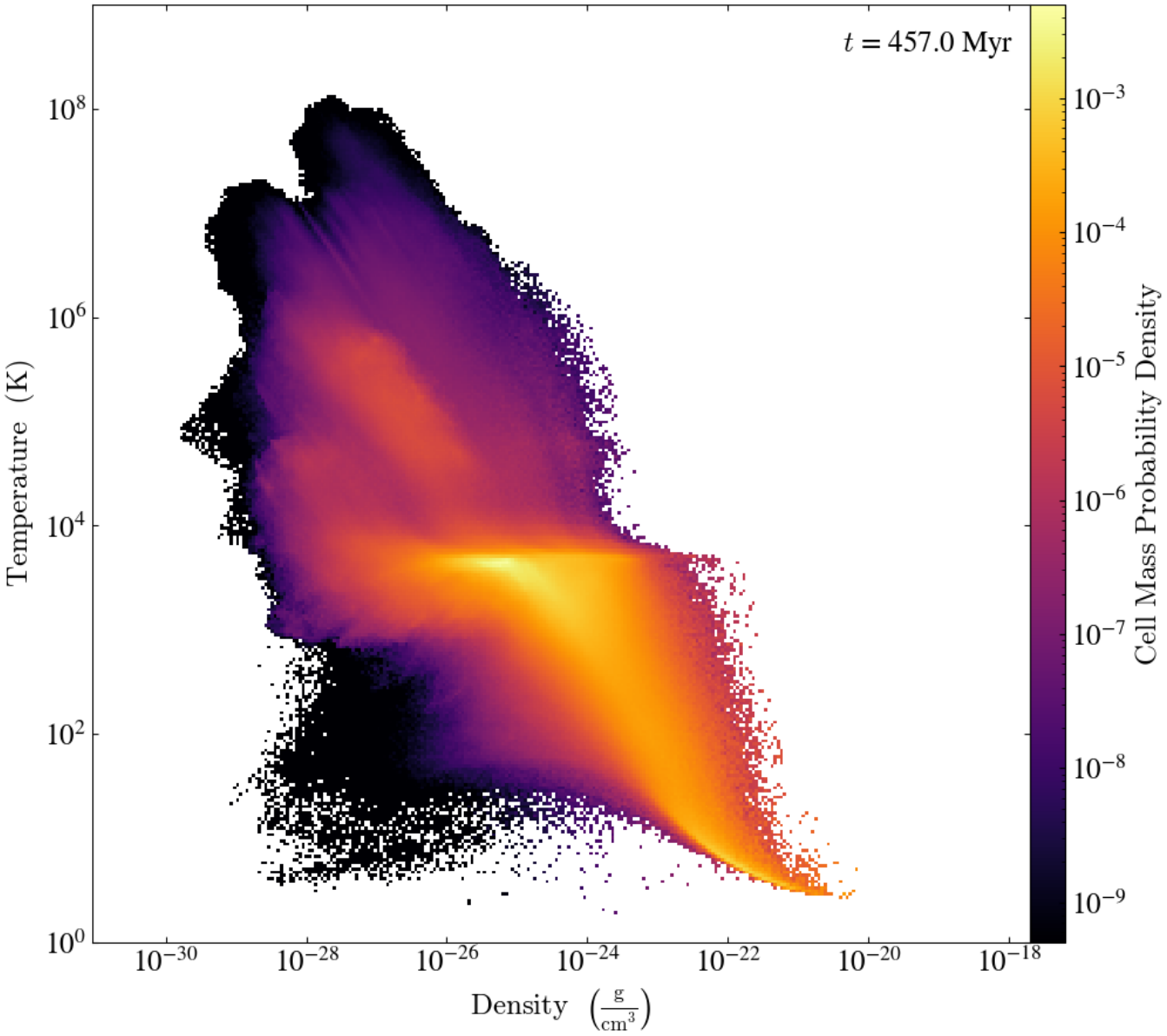}
\caption{
Probability distribution function of gas in the density--temperature plane {from} an isolated dwarf galaxy simulation performed with \gamer.
{An animation is available that spans $0\textrm{ -- }750\Myr$ with a real-time duration of 47 seconds. After $\sim 100\Myr$, the interstellar medium reaches thermal balance against radiative cooling, becoming dominated by diffuse warm gas with density $\sim 10^{-25}\,\rm{g\,cm^{-3}}$ and temperature $\sim 10^{4}\,\rm{K}$.}
}
\label{fig:dwarf-galaxy_phase}
\end{figure}

We simulate the evolution of an isolated dwarf galaxy analogous to the Wolf--Lundmark--Melotte (WLM) galaxy. The simulation is performed with \gamer, employing Hancock's MUSCL scheme, the HLLC Riemann solver \citep{Toro2009}, and the PPM data reconstruction. The initial conditions are constructed using \texttt{MakeDiskGalaxy} \citep{Springel2005} and then converted from particle data to AMR grids using \yt. The model consists of a dark matter halo with a mass of $M_{\rm halo} = 1\times10^{10}\Msun$, a gaseous disk of $M_{\rm gas} = 7\times10^7\Msun$, and a stellar disk of $M_{\rm disk} = 1\times10^7\Msun$. The grid refinement follows a quasi-Lagrangian criterion, where a cell is refined when its mass exceeds $10\Msun$. The simulation domain spans $450\kpc$, resolved with a $256^3$ root grid and nine refinement levels, achieving a finest spatial resolution of $3.4\pc$.

As the system evolves, the gas collapses under self-gravity and cools via metal-line emission calculated with \texttt{Grackle} \citep{Smith2017}. When the Jeans length can no longer be resolved by a single cell at the highest refinement level, a star particle forms stochastically, with a stellar mass resolution of $10\Msun$. Once a star particle reaches an age of $10\Myr$, it undergoes a core-collapse supernova, injecting $10^{51}\,\rm erg$ of thermal energy into the surrounding gas. This supernova feedback drives bubble-like blast waves that expand outward, thereby redistributing the gas and regulating star formation activity within the galaxy.

The simulation runs for $750\Myr$, with snapshots output every $10\Myr$. Each snapshot has a size of $\sim 6\GB$, resulting in a total storage requirement of about $450\GB$. We perform in situ analysis of both gas and particle data with \libyt\ every $1\Myr$. Specifically, \libyt\ first creates a \texttt{box} data object with a width of $4.5\kpc$ and plots the projected gas density within this region using \texttt{ProjectionPlot}. It then defines star and supernova particle types through \texttt{add\_particle\_filter} and annotates these particles on the projection plot using \texttt{annotate\_particles}. Finally, it generates a mass-weighted phase diagram of gas temperature versus density within the box using \texttt{PhasePlot}. See \fref{fig:dwarf-galaxy_slice} and \fref{fig:dwarf-galaxy_phase} for an illustration. This in situ analysis enables monitoring of the simulation at ten times higher temporal resolution compared to post-processing, while requiring negligible additional storage. Similar to the CCSN simulations, \libyt\ consumes only $\sim 3\%$ of the total simulation time.

\subsection{Fuzzy Dark Matter} \label{subsec:fdm}

\begin{figure}[ht]
\includegraphics[width=1.0\columnwidth]{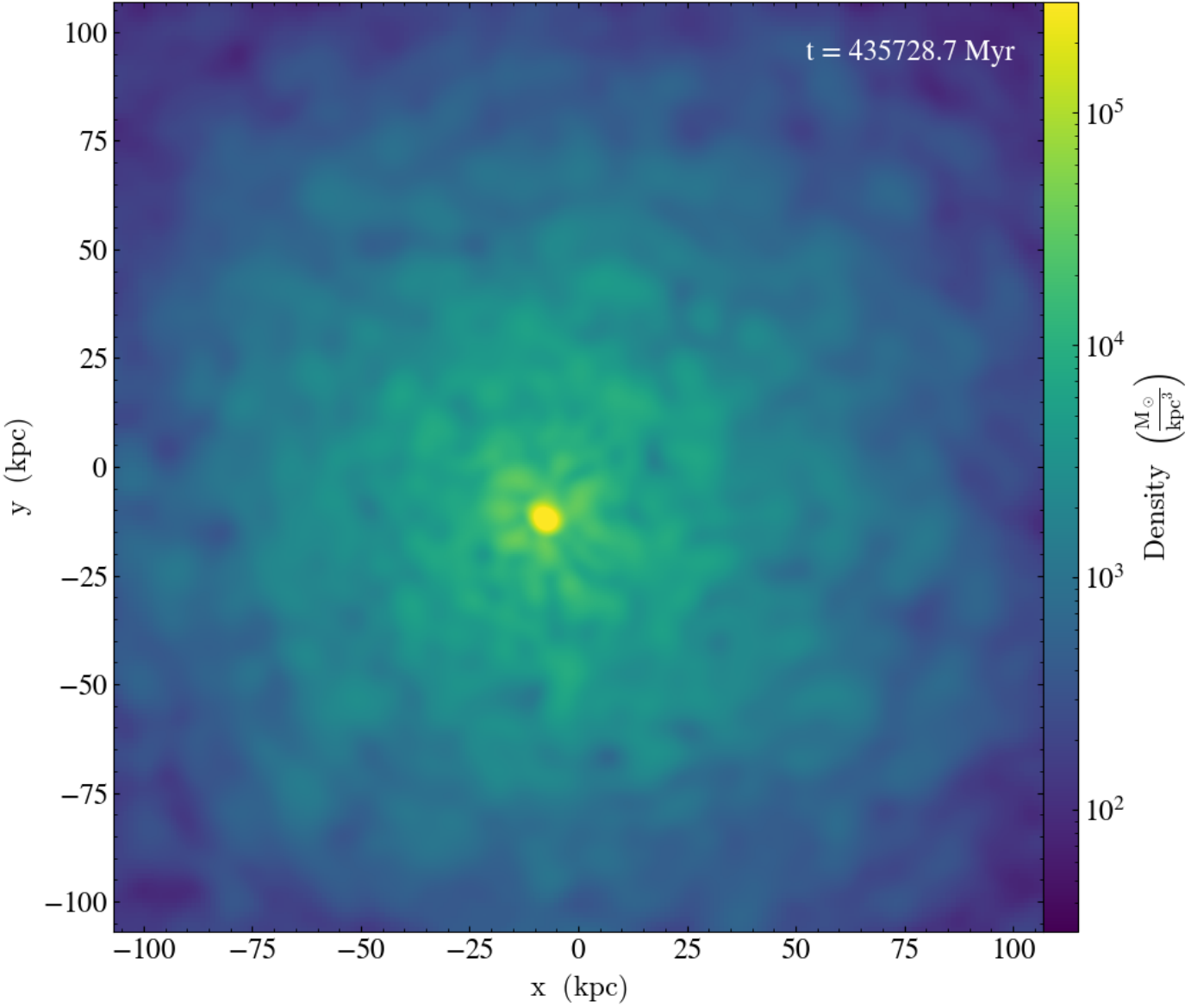}
\caption{
Average density distribution in an FDM simulation performed with \gamer. The yellow region marks a soliton core that forms spontaneously at the center of an FDM halo.
{An animation of this figure is available. It has a total duration of 113 seconds and illustrates the evolution from the initial condition to the onset of soliton condensation. Initially, the halo remains near equilibrium, exhibiting only granular density fluctuations. After $\sim 100\Gyr$, a soliton core with radius $\sim 3\kpc$ forms and gradually grows more compact and massive.}
}
\label{fig:fdm}
\end{figure}

We use \gamer\ to simulate the process of soliton condensation \citep{Levkov2018} in an FDM halo. The code solves the Schr\"odinger--Poisson equations using the local pseudo-spectral method of \citet{Kunkel2025}. The initial condition represents an isolated dark matter halo in equilibrium without a soliton core, self-consistently constructed using the eigenmode method of \citet{Lin2018}. The halo has a virial mass of $M_{\rm halo} = 2\times10^{10}\Msun$ and an FDM particle mass of $m_{\rm FDM} = 2\times10^{-23}\eV$. The simulation domain spans $428\kpc$, resolved with a $96^3$ root grid and four refinement levels, achieving a finest spatial resolution of $0.28\kpc$. Initially, the halo remains in equilibrium for an extended period, exhibiting only granular density fluctuations. After approximately $100\Gyr$ of evolution, a soliton core of radius $\sim 3\kpc$ spontaneously forms and gradually becomes more compact and massive over time.

The simulation runs for $435\Gyr$, with snapshots output every $10\Gyr$. Given the large uncertainty in the timescale of soliton condensation within an FDM halo, it is crucial to monitor this process at much higher temporal resolution. To this end, we use \libyt\ to plot the average line-of-sight FDM density every $32\Myr$ (see \fref{fig:fdm}), achieving a temporal resolution more than 300 times higher than that obtained from post-processing. This enables us to resolve the wave oscillation timescale over the entire condensation period while requiring negligible additional storage. In this simulation, \libyt\ accounts for $\sim 15\%$ of the total simulation time.

\subsection{Sod Shock Tube} \label{enzo:sod-shock-tube}

\begin{figure}[ht]
\includegraphics[width=1.0\columnwidth]{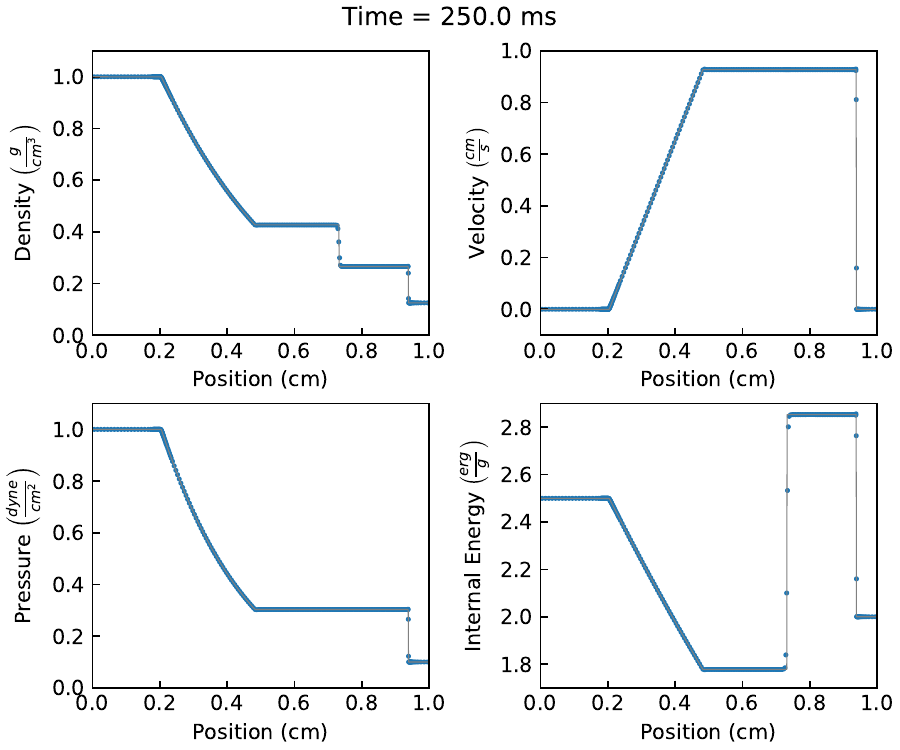}
\caption{
The classical Sod shock tube test results in density, velocity, pressure, and internal energy in \enzo\ simulation at $t=250\ms$.
The blue dots are results from the hydrodynamics solver in \enzo. 
The gray line represents the exact solution at $t=250\ms$, which is plotted in the static figure for comparison purposes only.
We use \yt\ \texttt{LinePlot} to sample the data points in the simulation, and then use \texttt{matplotlib} to plot the figure.
{An animation is available that spans the full simulation and samples the evolution every 5 ms in simulation time. All the results in chronological time steps are combined into a 3-second animation.}
}
\label{fig:sod-shock-tube}
\end{figure}

The Sod shock tube test \citep{SOD19781} is a classic problem for testing the Riemann solver and is included as an example in \enzo.
The test is small and can be performed with minimal computing resources using a single process.
Here, we perform the test solely to demonstrate that \libyt\ is applicable to one-dimensional AMR simulations and can operate in serial mode.

In the simulation, the governing intrinsic fields are density, velocity, and total energy.
We aim to monitor the evolution of density, velocity, pressure, and internal energy.
We use \yt\ built-in derived field to get pressure data, and we add our custom field, the internal energy, using \yt\ \texttt{add\_field}, by subtracting the square of velocity from total energy.
The data points are sampled using \yt\ \texttt{LinePlot}.
Next, we plot the data points using \texttt{matplotlib}.
The data pipeline is wrapped in a Python function, and \libyt\ calls this routine once every $5$ ms.
A total of $51$ time steps are processed, including the initial condition.
Instead of storing each simulation snapshot to disk and performing post-processing in an embarrassingly parallel manner, this approach only stores the generated figure at each time step.
We compare the simulation data (blue dots) at $t=250\ms$ to the exact solution (gray line), as shown in \fref{fig:sod-shock-tube}.

Despite the simulation being comparatively small and able to be performed by storing all datasets and then embarrassingly parallelizing the data plotting, merging the data processing workflows into the simulation reduces repetitive work. 
The post-processing script and the inline script are nearly identical, except for the way data is imported.
\libyt\ is not restricted to only \yt's functionality.
We can also combine other Python packages, such as \texttt{matplotlib}, here. 

\subsection{Kelvin-Helmholtz Instability} \label{enzo:kelvinhelmholtz}

\begin{figure}[ht]
\includegraphics[width=\columnwidth]{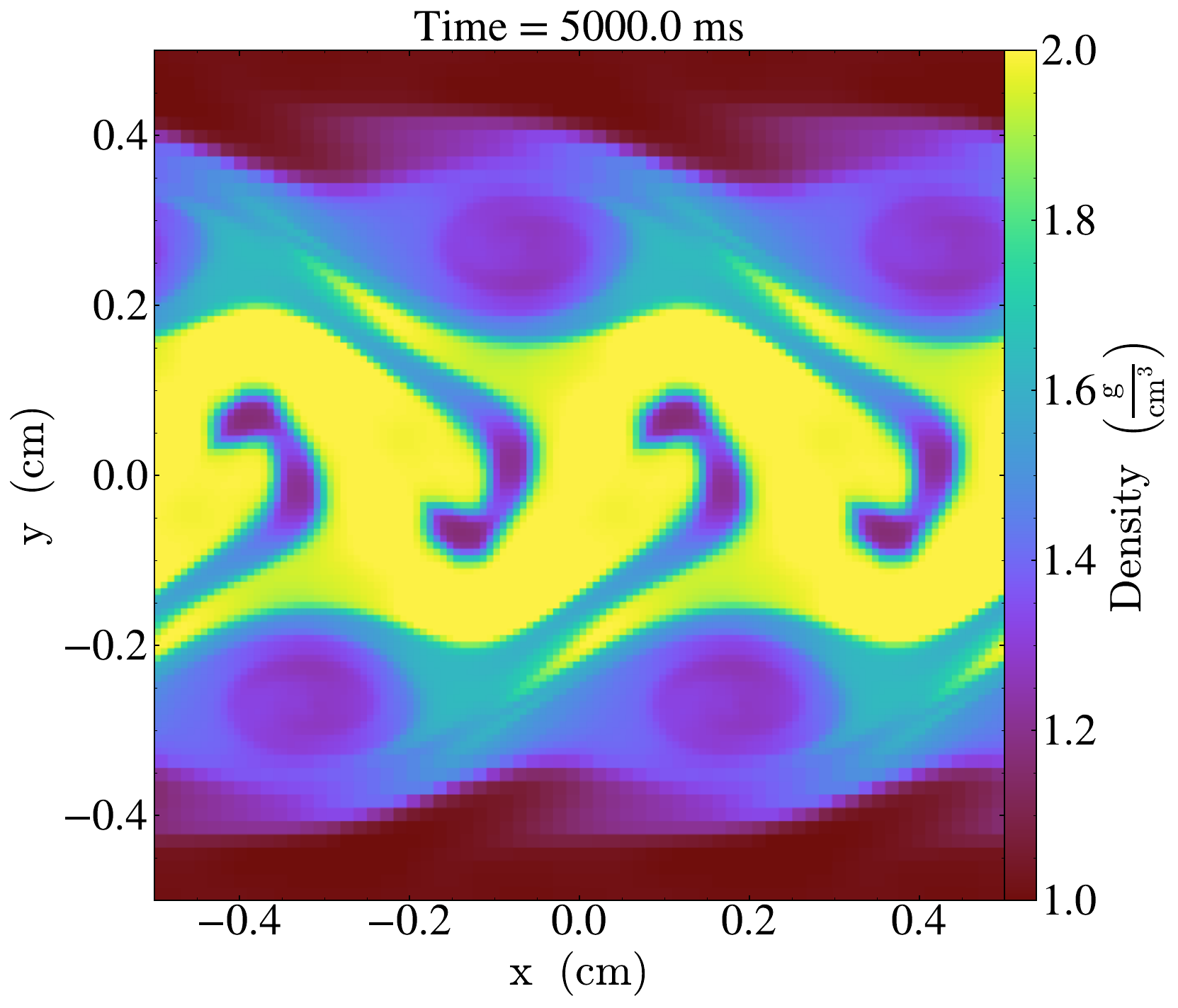}
\caption{
Density field from the Kelvin-Helmholtz instability test problem at $t=5000\ms$, obtained with \enzo.
We use \yt\ \texttt{SlicePlot} to plot the data.
{The available animation spans the full simulation and samples the evolution every 5 ms in simulation's physical time.}
All the plotted figures are combined into a 50-second animation.
{It closely tracks the simulation from the initial condition to the turbulent state.}
}
\label{fig:kelvin-helmholtz}
\end{figure}

\enzo\ includes a Kelvin-Helmholtz instability test problem, in which two fluids flow in opposite directions on a periodic two-dimensional grid.
The simulation can be performed serially on a single machine.
This demonstrates the functionality of \libyt\ in processing a two-dimensional simulation.

The intrinsic fields in this test are density, x-velocity, y-velocity, and total energy.
We use \yt\ \texttt{SlicePlot} to visualize the turbulent flow every $5\ms$.
A total of $1001$ time steps are analyzed.
\fref{fig:kelvin-helmholtz} shows the density field at $t=5000\ms$.
By invoking the inline Python function at each time step, \libyt\ integrates the entire data-processing pipeline into the simulation workflow, thereby eliminating repetitive post-processing steps.

\subsection{AGORA Galaxy} \label{enzo:agora-galaxy}

The AGORA Project \citep[][\url{https://sites.google.com/site/santacruzcomparisonproject/}]{Agora} is dedicated to investigating galaxy formation using different high-resolution simulations and comparing their results.
\enzo\ is part of the project.
{This section presents the AGORA galaxy simulation using \libyt\ to closely monitor the complete evolution with high temporal resolution, without truncating the spatial resolution or corrupting disk space.}

The simulation is performed on a local high-performance computing cluster, utilizing $4$ computing nodes, each of which hosts one MPI process.
The simulation physics setup includes radiative feedback and star particle creation, but without star particle feedback. 
The simulation domain spans $1.311\Mpc$ and is covered by a root-level grid with eight refinement levels.
It has a spatial resolution $80$ pc at the maximum refinement level.
Our target region is a box with a width of $30\kpc$ at the center of the simulation domain, corresponding to roughly $2\%$ of the domain's width.
We aim to track the evolution of galaxy formation, including density, temperature, and particle distribution, within this comparatively small region.

\begin{figure*}[ht]
\centering
\subfigure{\includegraphics[width=0.315\textwidth]{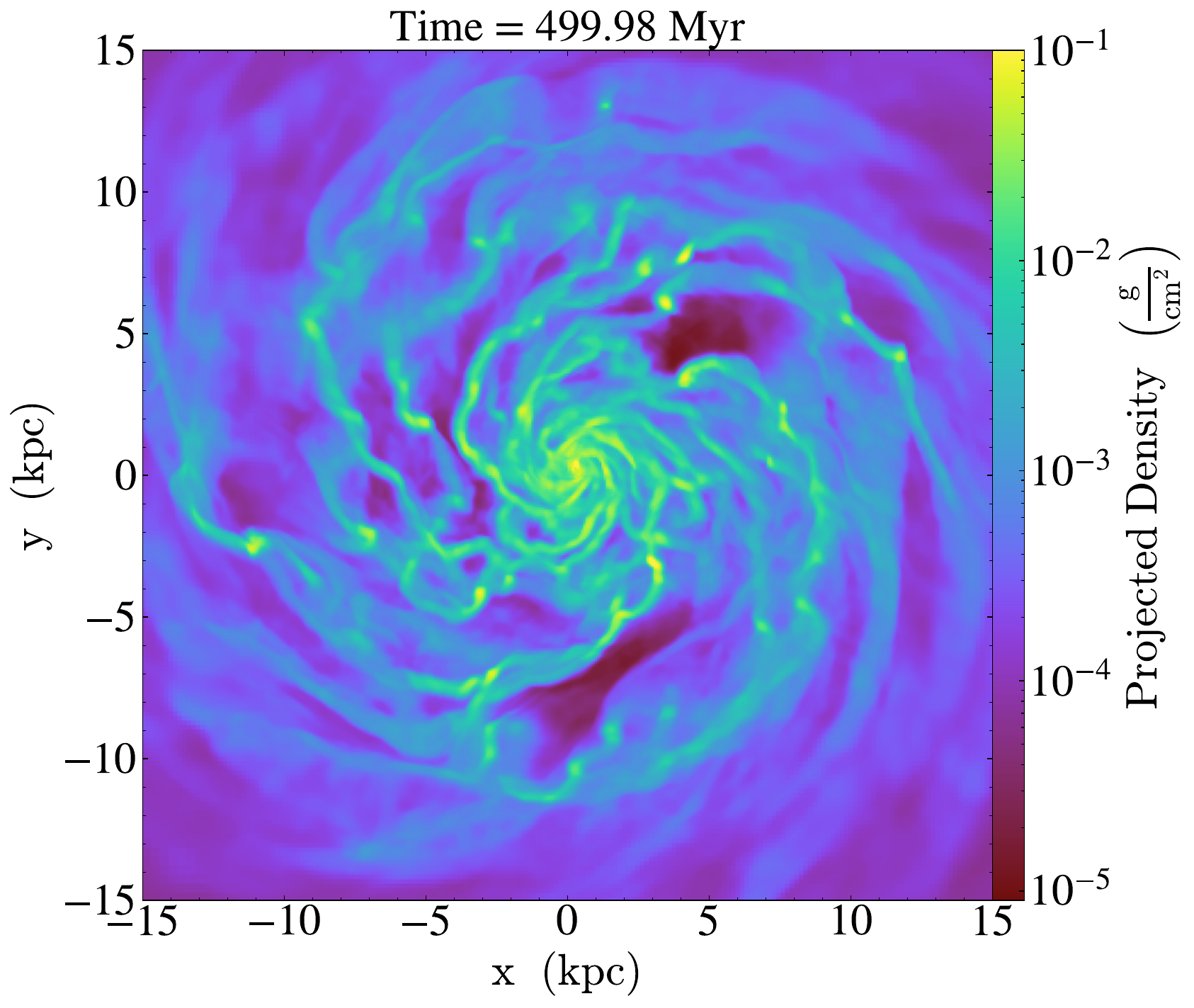}}
\subfigure{\includegraphics[width=0.305\textwidth]{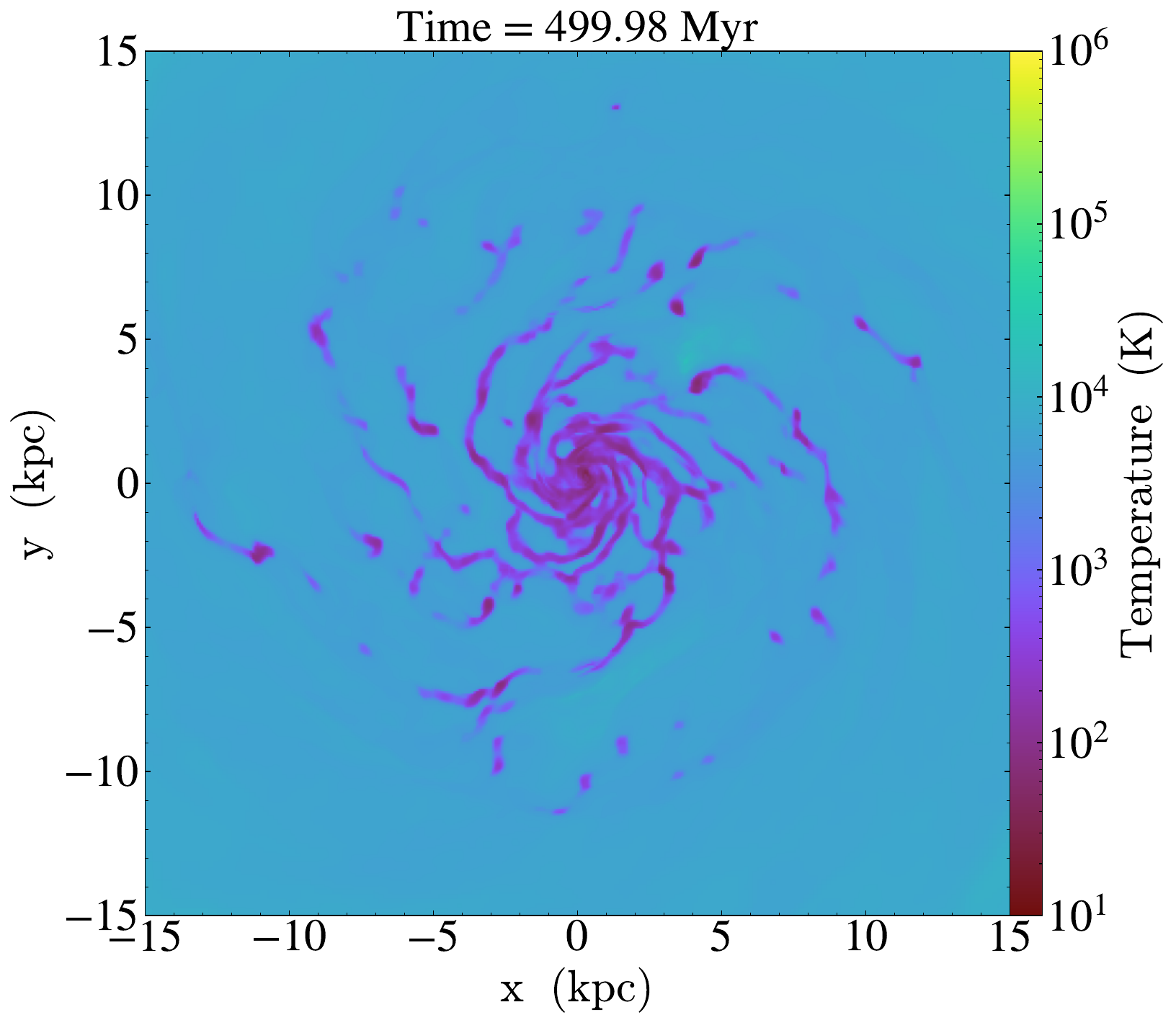}}
\subfigure{\includegraphics[width=0.305\textwidth]{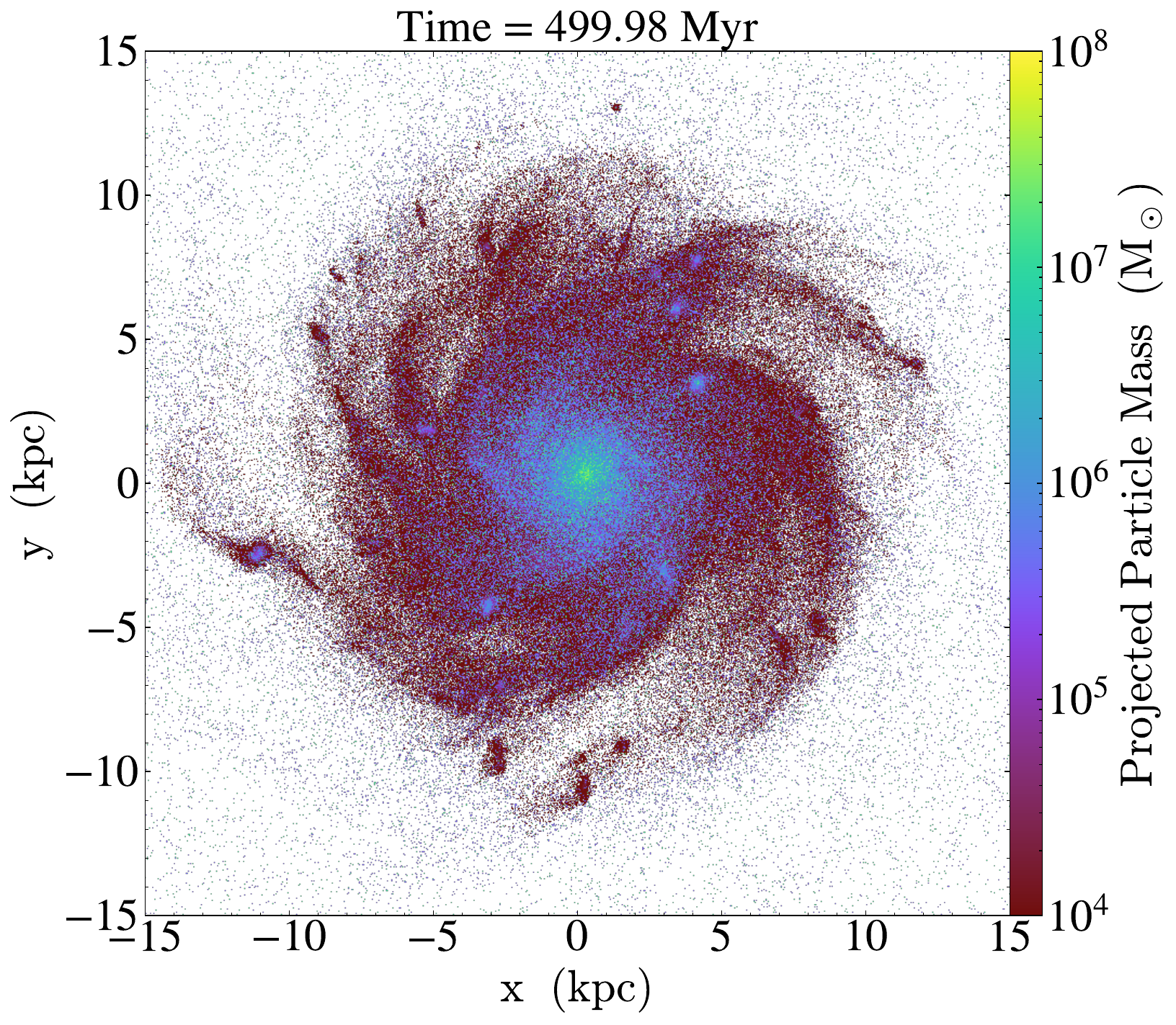}}
\subfigure{\includegraphics[width=0.315\textwidth]{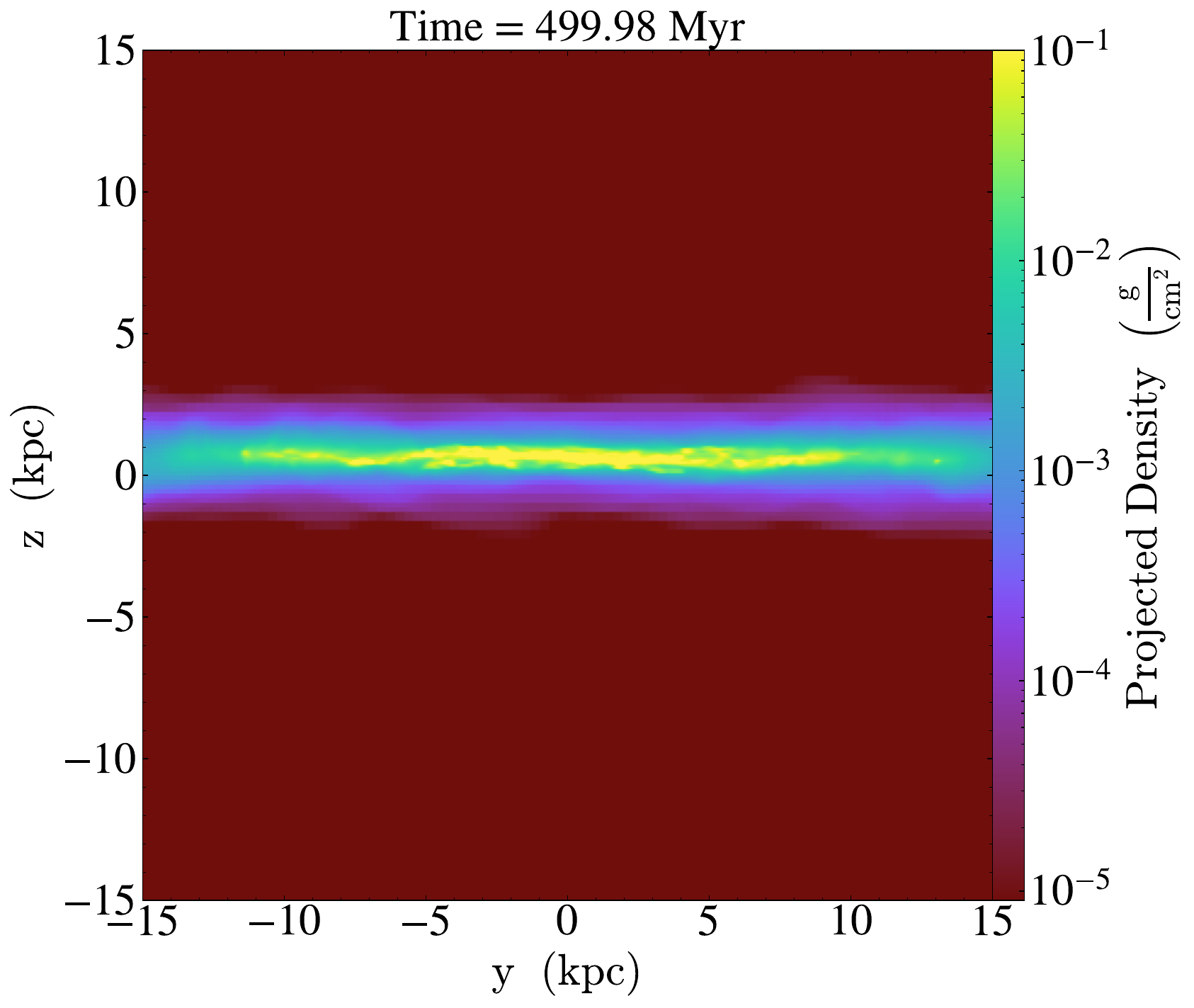}}
\subfigure{\includegraphics[width=0.305\textwidth]{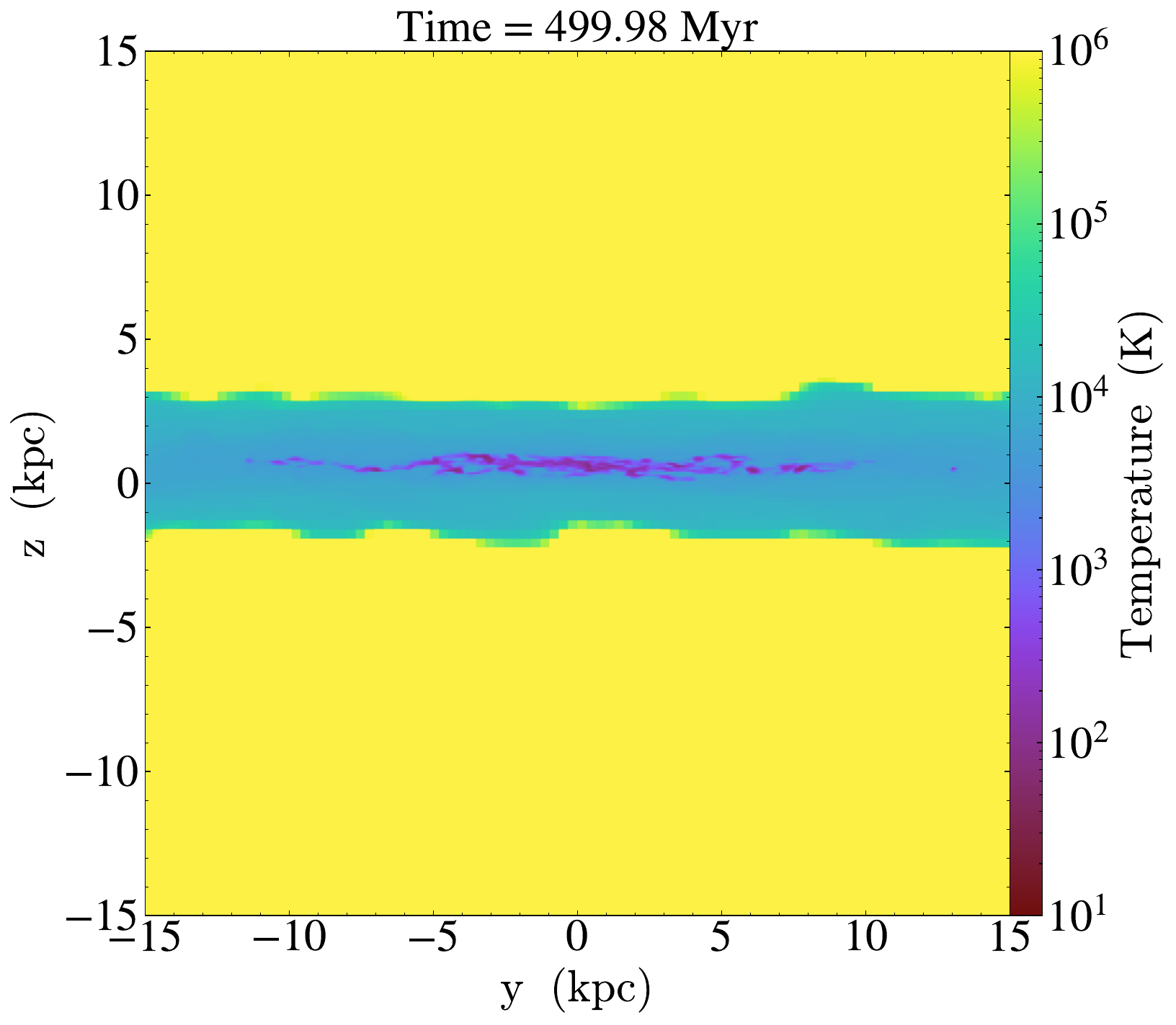}}
\subfigure{\includegraphics[width=0.305\textwidth]{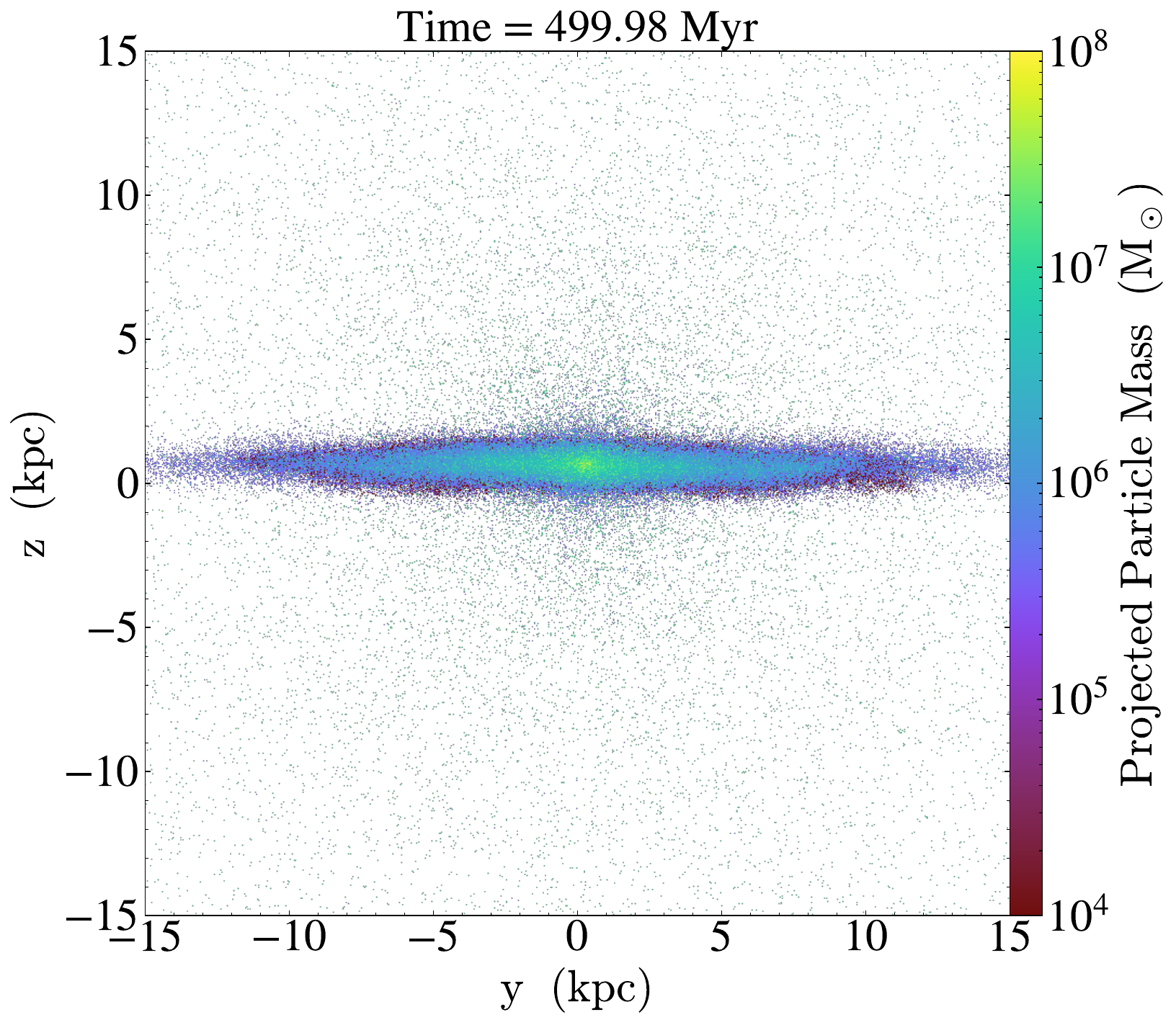}}
\caption{
{Final snapshot ($t=499.98\Myr$) of the AGORA galaxy simulation using \enzo, without stellar feedback.
Each panel is centered on the simulation domain and covers a width of $30$ kpc. 
Columns display density (left), density-squared-weighted temperature (center), and particle distributions annotated with dark matter clump and star particle masses (right), including those formed during the simulation and those present in the initial condition. 
The top and bottom rows show face-on and edge-on projections, respectively.
We closely monitor the dynamics of galaxy formation by calling \libyt\ Python data analysis routine every 0.5 Myr. 
The resulting figures are then combined into a 50-second animation that shows the sequence from 0 to 499.49 Myr.
The animation illustrates the evolution from the initial gas and dark matter distribution to the formation of a rotationally supported disk galaxy with spiral structure.
This approach enables high-temporal-resolution monitoring of galaxy formation without compromising spatial resolution or requiring substantial disk storage.
}
\label{fig:agora-galaxy}}
\end{figure*}

Each simulation snapshot is $\sim2$ GB.
We want to visualize and track the evolution every $0.5\Myr$ from $0$ to $500\Myr$, which contains $1{,}001$ time steps, including the initial condition.
Thus, for a traditional post-processing method, approximately $2{,}000$ GB of hard disk space is required, and then we will need to embarrassingly parallelize the data processing and produce the final results using \yt\ \texttt{ProjectionPlot} and \texttt{ParticlePlot}.
However, we do not necessarily need the data outside of our target region, and we can merge the original post-processing routine into the simulation runtime by invoking the Python function using \libyt.
This eliminates the intermediate step of writing data snapshots to disk, resulting in a significant reduction in disk space usage.
{The Python script includes operations like defining new fields, annotating titles, object selections, setting z-axis limitations.}
Because converting the post-processing script to the inline script is a two-line change (\sref{subsec:connecting-libyt-to-yt}),
there is no barrier, and it is easy to apply our method to an already written Python script initially targeting the post-processing process.
{\fref{fig:agora-galaxy} shows projections of the density field (left column), the temperature field weighted by the square of the density (center column), and particle distributions (right column) annotated with dark matter clump and star particle masses, including both particles formed during the simulation and those present in the initial conditions, at the final time step $t=499.98\Myr$. 
The upper and lower rows correspond to face-on and edge-on projections, respectively.}
{These figures aim to replicate the results of the \enzo\ simulation in the AGORA Project, while allowing galaxy formation to be tracked in great detail without sacrificing spatial resolution or incurring significant disk storage costs.}

\section{Discussion} \label{sec:discussion}

\subsection{Open Source Software and Maintenance} \label{subsec:open-source-software}
\libyt\ is free and open source; it does not depend on non-free or non-open source software. 
The code is maintained on GitHub.
\tref{tab:open-source-software} lists the important online resources
of \libyt.

\begin{deluxetable}{ll}[htb]
\tablewidth{\columnwidth}
\tablecaption{A List of Resources for \libyt\label{tab:open-source-software}}
\tablehead{
\colhead{Source} & \colhead{Link}
}
\startdata
    \libyt\   & \url{https://github.com/yt-project/libyt} \\
    \texttt{yt\_libyt} & \url{https://github.com/data-exp-lab/yt_libyt} \\
    \texttt{jupyter\_libyt} & \url{https://github.com/yt-project/jupyter_libyt} \\
    Documentation & \url{https://libyt.readthedocs.io/} \\
    Slack channel & \url{https://yt-project.slack.com/channels/libyt} \\
    Mailing list  & \href{mailto:libyt-users@googlegroups.com}{libyt-users@googlegroups.com}
\enddata
\end{deluxetable}

We use GitHub Actions to automate testing and profiling for new pushes and pull requests.
They cover unit testing, memory profiling, building and installing using different options, and end-to-end testing.
We write the unit test suites using the \texttt{googletest} framework. 
The unit tests contain binding AMR data structures, supporting Python shell, instantiating NumPy arrays, and MPI communications, including big MPI, as well as one-sided communication for the data redistribution process.
These tests are built and run with or without MPI in combination with different Python versions ranging from $3.7 \sim 3.14$.
We test against two MPI implementations: \texttt{openmpi} \citep[][\url{https://www.open-mpi.org/}]{openmpi} and \texttt{mpich} (\url{https://www.mpich.org/}).
The code coverage is collected and uploaded to Codecov\footnote{\url{https://about.codecov.io/}}.
We perform memory profiling using \texttt{valgrind} \citep[][\url{https://valgrind.org/}]{valgrind}.
We profile all combinations of building \libyt\ from either the pure Python C API or \texttt{pybind11}, with or without MPI.
The resulting profiles are uploaded as artifacts in GitHub Actions.
We also test whether the build options, when toggled on or off, can successfully build and install to a prefix location on Linux and macOS operating systems.
We run an end-to-end test in parallel (MPI) and serial (without MPI) modes, which consists of a mock simulation with \libyt\ implemented and an in situ analysis script using \yt\ to run operations like projections, slice plots, and particle plots using intrinsic and derived data,
ensuring that \libyt\ can correctly call \yt\ functionalities during in situ analysis.

We use the code-formatting tools \texttt{clang-format}\footnote{\url{https://clang.llvm.org/docs/ClangFormat.html}} and \texttt{cmake-format}\footnote{\url{https://cmake-format.readthedocs.io/}} to govern the coding style of source code and CMake files.
They are integrated into \texttt{pre-commit} to ensure code is always formatted before committing to \texttt{git} history.
We use \texttt{doxygen}\footnote{\url{https://www.doxygen.nl/}} to generate inline code documentation.

Together, these tools help us maintain the source code of our in situ analysis packages more effectively and ensure that it remains consistent with other well-designed tools.

\subsection{Time Performance} \label{subsec:time-performance}

\subsubsection{Weak Scaling and Strong Scaling of Libyt} \label{subsubsec:weak-strong-scaling-libyt}

{
We use the \libyt\ data redistribution process (\sref{subsec:parallelism}) as the basis for these scaling tests because it represents a common and performance-critical workflow in in situ analysis.
In practice, user-defined Python analysis uses the \libyt\ API to read data from the local process or fetch data from other processes, requiring all MPI processes to enter the data-reading phase collectively.
This introduces synchronization and communication overhead that may limit scalability at large processor counts.
By evaluating the redistribution of randomly selected grids to randomly selected processes, we quantify the cost of these collective operations and characterize the impact of \libyt\ data redistribution on performance as the system scales.
}

\begin{figure}[ht]
\centering
\includegraphics[width=0.49\textwidth]{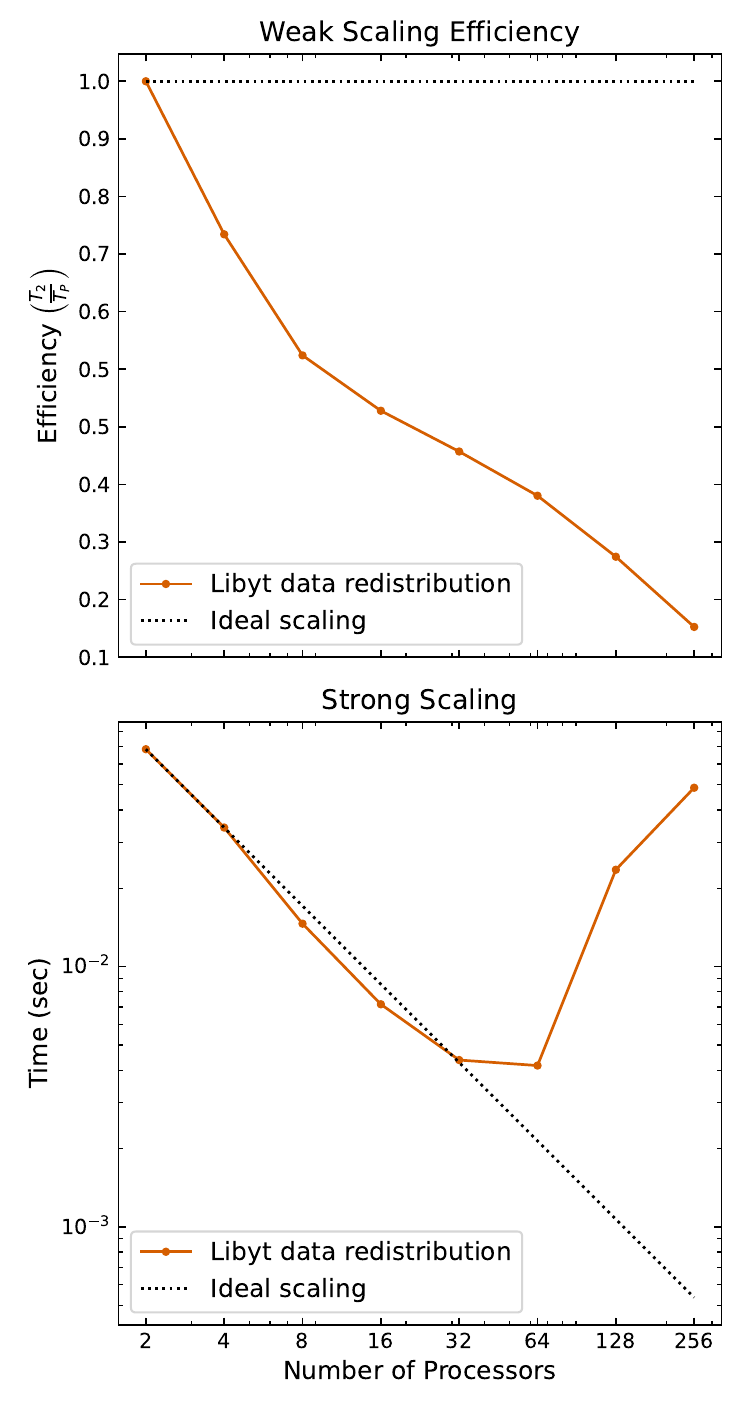}
\caption{
{Weak scaling efficiency (upper) and strong scaling (lower) of the \libyt\ data redistribution process.
Each MPI process runs on a single core and occupies a computing node (two processes per node at $256$ processors).
The tests redistribute randomly selected $8^3$ floating-point grids among processes.
In weak scaling, each process fetches $5{,}000$ grids ($\sim 10\MB$), maintaining a constant workload per process; 
efficiency declines approximately linearly with $\log($processor count$)$ due to the synchronization overhead.
In strong scaling, a fixed $20{,}000$ grids ($\sim 40\MB$) are redistributed evenly among processes; performance follows the ideal trend up to $32$ processors, after which execution time increases as communication overhead dominates, partly because the reduced computational workload per process becomes insufficient to offset the communication cost.}
\label{fig:libyt-weak-strong-scaling}}
\end{figure}

{
In both the weak and strong scaling tests, each MPI process runs on a single CPU core and occupies an entire computing node.
When the number of processors reaches $256$, two MPI processes share a node due to resource constraints.
The tests focus on redistributing randomly selected grids of $8^3$ floating-point cells among random processes.
In the weak scaling test, each MPI process fetches $5{,}000$ grids ($\sim 10\MB$).
As the number of processors increases, the total number of grids redistributed by \libyt\ grows proportionally, maintaining a constant workload per process.
In the strong scaling test, the total number of grids redistributed by \libyt\ is fixed at $20{,}000$ ($\sim 40\MB$). 
These grids are evenly divided among all processes, so the number of grids fetched from other processes per process decreases as the total number of processors increases.
}

{
All data redistribution procedures are implemented within a Python function. 
We measure the execution time from the start of the function to its completion and record the maximum time across all MPI processes.
The test is repeated for five iterations, and the reported time is the average over all iterations except the first, which is omitted to avoid initialization effects.
The weak scaling efficiency is defined as the execution time on $2$ processors ($T_2$) divided by the execution time at each processor count ($T_P$).
\fref{fig:libyt-weak-strong-scaling} shows the weak scaling efficiency (upper panel) and strong scaling performance (lower panel) of the \libyt\ data redistribution process.
}

{
The approximately linear decline in weak scaling efficiency with respect to $\log($processor count$)$ reflects the growing overhead associated with collective operations in the \libyt\ data redistribution process.
The strong scaling results follow the ideal trend up to approximately $32$ processors, beyond which the execution time increases dramatically, indicating that communication overhead begins to dominate.
This behavior may also result from the relatively small number of redistributed grids handled per process at large processor counts, leaving the reduced computational workload insufficient to offset the communication costs.
The collective design of the \libyt\ data redistribution enables flexible execution of arbitrary Python scripts; however, this trade-off introduces synchronization overhead that can limit scalability.
Future work may explore asynchronous coordination strategies to mitigate global synchronization costs.
}

\subsubsection{Comparing Libyt to Post-Processing} \label{subsubsec:time-compare-to-postprocessing}

\begin{figure}[ht]
\centering
\includegraphics[width=\columnwidth]{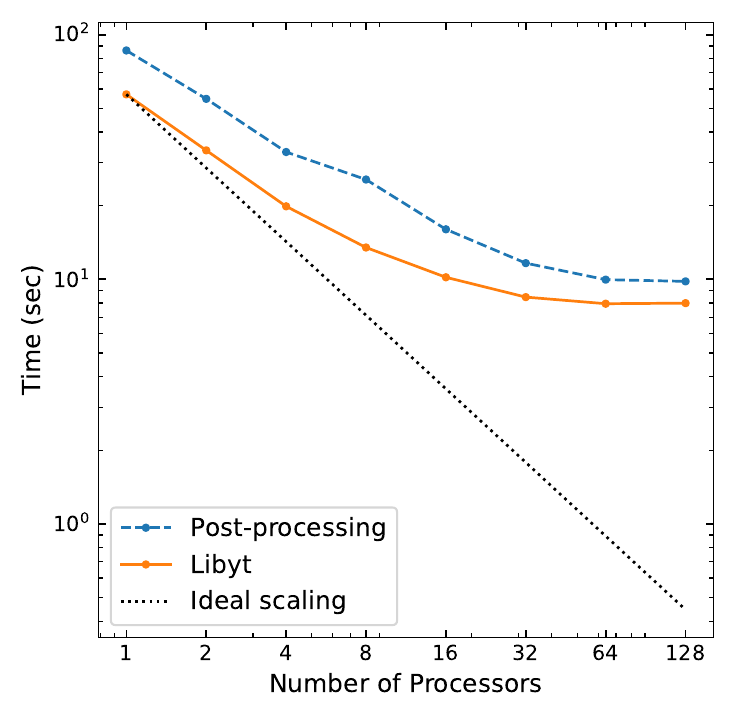}
\caption{
Strong scaling of \libyt. 
Similar tests were done in {our previous papers \citep{libyt-scipy2023, libyt-pasc-2024}}.
Here, we re-measure performance using the same test problem but with more recent versions of \libyt\ and \yt.
The test compares the performance between in situ analysis
with \libyt\ and post-processing for computing 2D profiles on a \gamer\ dataset
with five AMR levels and a total of $9.0 \times 10^8$ cells.
It measures the total execution time of the Python {analysis function from invocation to result generation in both methods.}
\libyt\ outperforms post-processing by $\sim 20 \textrm{ -- } 45\%$
since the former does not require loading data from disk.
The dotted line shows the ideal scaling for reference.
\libyt\ and post-processing show a similar deviation from the ideal scaling
when increasing the number of processes, as expected, since \libyt\ directly utilizes the MPI parallelism of \yt\ and therefore the scaling bottleneck lies in \yt.
\label{fig:time-proc-ideal}}
\end{figure}

{In our previous work \citep{libyt-scipy2023, libyt-pasc-2024}, we evaluated the strong scaling performance of \libyt\ and compared it with post-processing in a real application.}
Here, we use more recent \libyt\ and \yt\ versions and re-measure the strong scaling performance shown in \fref{fig:time-proc-ideal}.

The time performance test uses an increasing number of processors and a fixed problem size.
We use \gamer\ (version $2.2.0$) to run the Plummer simulation test problem, with and without \libyt\ (version $0.2.0$), for six iterations, and use \yt\ (version $4.3.1$) to compute the 2D profile.
We then take the average performance, excluding the initial condition setup.
Both the post-processing method and \libyt\ compute the 2D profiles from the same time-series data. 
The former operates on datasets stored on the hard disk, whereas the latter operates directly on the distributed simulation data in memory.
The dataset consists of 3,430 grids, with a $640^3$ root-level grid and four refinement levels, each containing $64^3$ cells, resulting in a total of $9.0 \times 10^8$ cells. 
We compare the time performance between post-processing and in situ analysis using \libyt\ with the same Python script (except for the line loading a dataset; see \sref{subsec:performing-in-situ-analysis}).
{We measure the execution time from invoking the Python analysis function to figure generation, excluding data binding and simulation runtime in \libyt, and excluding the data snapshot output time required for post-processing.}
\libyt\ outperforms post-processing by $\sim 20\textrm{ -- }45\%$ since the former avoids loading data from disk to memory.
{Because \libyt\ directly leverages the MPI parallelism of \yt\ (see \sref{subsec:parallelism}), the parallel efficiencies of \libyt\ and post-processing are similar, and both deviate from the ideal scaling as the number of processors increases.}
{However, at large processor counts, the cost of \libyt\ data redistribution (\sref{subsubsec:weak-strong-scaling-libyt}) becomes more significant and is insufficient to offset the communication overhead.
Consequently, \libyt's performance degrades and approaches that of post-processing.}

The speed-up reaches a limit imposed by the serial fraction of the workload, in accordance with Amdahl's law.
This saturation arises from components that cannot be parallelized, such as {Python imports}, the indexing data structure currently used in \yt, and the partitioning of dispatched jobs across processes. 
\yt\ stores and sorts AMR grid information (e.g., grid left/right {edges}, levels) in a linear format.
Grid information is stored in arrays with lengths equal to the number of grids.
This causes low performance and large memory consumption for simulation codes like \gamer, which divide the domain into many small grid patches and therefore produce a large number of grids. 
\gamer\ improves the performance by combining the eight nearby grid patches into 
a patch group or by using larger grid patches when appropriate, {thereby reducing the total number of grids.}
When writing datasets to disk, \gamer\ treats each patch group as a single logical \textit{grid}. 
This technique is also applied when measuring the time performance using \libyt\ derived field functionality (see Section \ref{subsubsec:field-information}).

\subsection{Memory Performance} \label{subsec:memory-performance}

\begin{figure*}[ht]
\centering
\includegraphics[width=\textwidth]{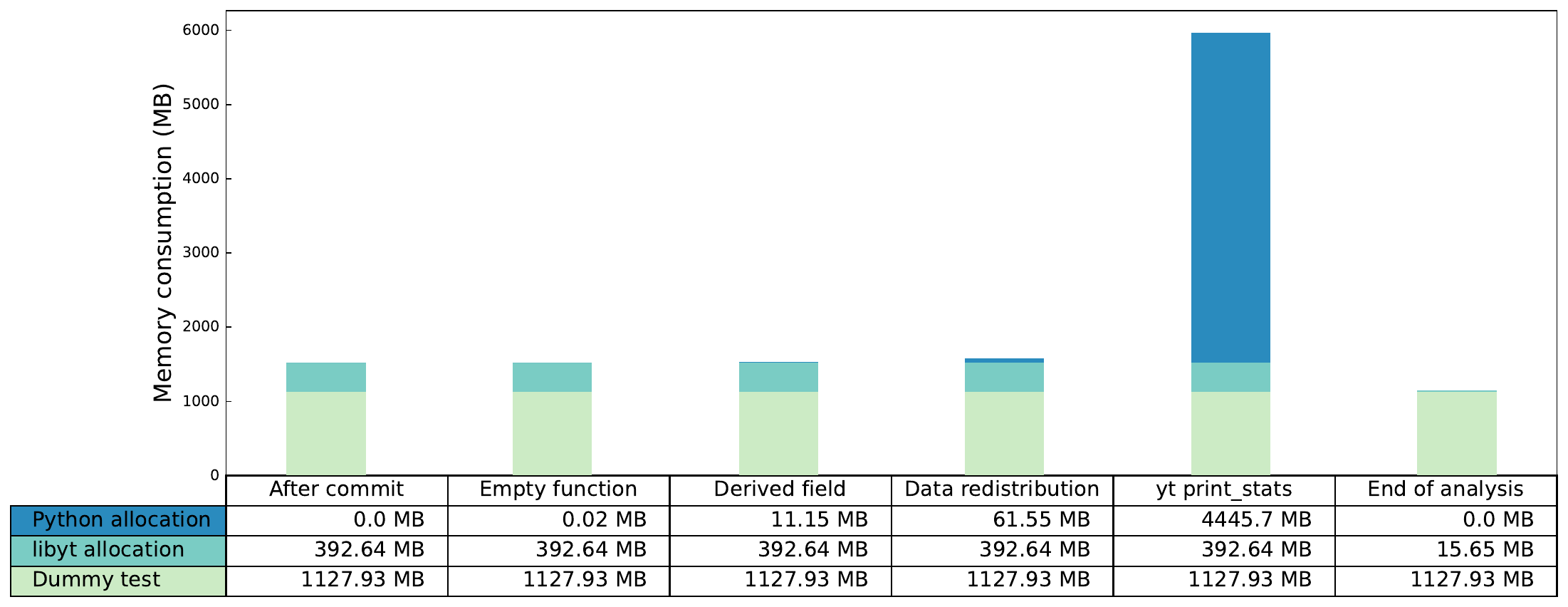}
\caption{
The average per-process memory usage across several \libyt\ in situ analysis stages, including: 
after committing the simulation settings (After commit), while running an empty Python function (Empty function), while generating $5,000$ grids of derived field data (Derived field), while redistributing $20,000$ grids (Data redistribution), run \yt\ operation \texttt{print\_stats} (yt \texttt{print\_stats}), and after freeing every resource allocated for this round of analysis (End of analysis).
The dummy test runs on 4 MPI processes and consists {of} a total of $2 \times 10^6$ grids, each containing one field of size $8^3$ with data type \texttt{float}.
\libyt\ takes $\sim 0.2$ KB to store a grid's metadata (e.g., grid level, field data pointers, etc).
When generating derived field data or performing data redistribution using \libyt, the memory consumption is controllable if data chunking is done wisely.
The Python memory allocation depends solely on the Python code executed, and \libyt\ has no control.
For example, \yt\ operation \texttt{print\_stats} takes four times more memory than the dummy test itself.
At the end of \libyt\ in situ analysis, because \libyt\ needs to store other background resources, it still occupies some memory and will not return to zero.
\label{fig:mem-prof}}
\end{figure*}

We build a dummy test consisting of $2 \times 10^6$ grids, all at the root level.
Each grid contains $8^3$ cells and has two fields: one cell-centered field with no ghost cells and one derived field.
We initialize this dummy cell-centered field with an arbitrary value of data type \texttt{float}.
The test uses 4 MPI processes, with the grids evenly distributed among them.
We use the \texttt{valgrind} profiling tool \texttt{massif}\footnote{\texttt{massif} is a heap profiler. It can also be used to profile stack memory using \texttt{--stacks=yes}.} to measure the memory consumption in different stages of \libyt\ in situ analysis. 
The details are described in Appendix \ref{appendix:memory-profile}.

\fref{fig:mem-prof} shows the memory allocation by dummy test, \libyt, and Python.
The memory profiles are sampled at: after committing simulation information (After commit), while running an empty Python function (Empty function), after generating 5,000 grids of derived field in Python (Derived function), after redistributing 20,000 grids with each process preparing 5,000 grids and fetching 15,000 from other processes in Python (Data redistribution), run \yt\ operation \texttt{print\_stats} (yt \texttt{print\_stats}), and the end of \libyt\ in situ analysis after freeing every prepared resource (End of analysis).
The tests are independent and measured in separate runs. 
The method for sampling memory consumption within Python functions is described in Appendix \ref{appendix:memory-profile}.
We calculate the average total memory consumption across all iterations and MPI processes at each sample point.
The number depicts the average memory consumption in an MPI process allocated by dummy test, \libyt, and Python at the sample points.
Dummy test represents the raw simulation memory consumption when running without \libyt.
\libyt\ allocation represents the resources allocated by \libyt\ in preparation for the in situ analysis, 
and it is calculated by subtracting the dummy test allocation from the total memory consumption after committing simulation information. 
Python allocation represents the additional resources designated while executing Python code, which is obtained by subtracting the \libyt\ and the dummy test allocations from the total memory consumption sampled after a specific Python operation has been completed inside the tested Python function.
The Python scripts used for memory profiling are provided in Appendix \ref{appendix:memory-profile-python-scripts}.

\libyt\ takes $\sim 0.2$ KB to store a grid's metadata.
Overall, the memory consumption of the back-communication methods in the \libyt\ Python Module, such as generating derived field data (Derived field) and data redistribution (Data redistribution), is manageable and controllable if data chunking is done wisely.
Memory allocation during the Python runtime depends on the Python code being executed.
It can range widely, from close to zero (Empty function) to several times larger than the original simulation size.
When running the \yt\ operation \texttt{print\_stats} (yt \texttt{print\_stats}), which involves traversing the indexing system that includes $2 \times 10^6$ of grids, 
it occupies four times more memory than the dummy test in a single process on average, though this is an extreme case.
Furthermore, because \texttt{print\_stats} handles the workload solely on the MPI root process, there is a spike in its memory consumption while the other processes remain unchanged.
Since the table represents the average memory consumption in a single process, the number is smoothed out.
At the end of in situ analysis and after freeing resources assigned for in situ analysis, \libyt\ still holds the Python interpreters and some runtime data (e.g., an inline Python script), so the memory consumption does not drop to zero.

From the test Python scripts, if the Python code requires large memory allocations or the job decomposition mechanism is not evenly distributed among the MPI processes, causing some processes to overload with jobs, \libyt\ has no control over it and cannot prevent Python from acquiring resources.
It is entirely determined by how Python scripts handle data analysis tasks.
This may lead to a runtime failure (e.g., a segmentation fault) if the memory consumption exceeds the hardware's limitations.
One solution is assigning more memory resources to each simulation process.
This can be either running a simulation using more MPI processes, so that each process holds less simulation data and has more memory space for Python, 
or simply running the simulation using machines with large memory.
Another solution is to separate the Python analysis process from the simulation process on designated computing nodes, so that the nodes don't need to hold both simulation and Python runtime data simultaneously. 
We will address this in future development.

{
Following the methodology described in this section and Appendix \ref{appendix:memory-profile}, we apply the same profiling procedure to real applications.
\tref{tab:app-mem-prof} presents the memory profiles of the \gamer\ core-collapse supernova simulation (\sref{subsec:ccsn}) and the \enzo\ AGORA galaxy simulation (\sref{enzo:agora-galaxy}), both with and without \libyt, under identical computing resource configurations.
For the \gamer\ case, memory profiles are captured at both the initial and intermediate stages and then averaged, since the simulation size increases over time.
All reported values are averaged over all sampled time steps and MPI processes.
In the table, ``Simulation'' denotes the memory consumption without \libyt, ``\libyt\ allocation'' denotes the additional memory allocated by \libyt\ to prepare for in situ analysis, and ``Python allocation'' denotes the extra memory consumed during Python runtime.
}

\begin{deluxetable}{lrr}[htb]
\tablewidth{\columnwidth}
\tablecaption{Memory Profile of \gamer\ and \enzo\label{tab:app-mem-prof}}
\tablehead{
 & \colhead{CCSN (\gamer)} & \colhead{AGORA galaxy (\enzo)}
}
\startdata
    Python allocation & 346.40\,MB & 146.58\,MB \\
    \libyt\ allocation & 248.14\,MB & 299.85\,MB \\
    Simulation        & 3132.84\,MB & 550.95\,MB \\
\enddata
\tablecomments{{Values represent the averaged memory consumption per MPI process.}}
\end{deluxetable}

{
In the \gamer\ case, per-process memory consumption increases by $8\%$ compared to running without \libyt, and by $19\%$ overall when including the preparatory steps needed to generate the slice plot.
}
{
In the \enzo\ case, instead of implementing derived field functionality (see Section \ref{subsubsec:field-information}) for quantities such as temperature and cooling time, these fields are generated in advance during the data-binding stage in \libyt.
As a result, the memory allocated by \libyt\ is higher, leading to a $54\%$ increase in per-process memory consumption compared to the baseline simulation without \libyt.
For demonstration purposes, we measure the memory consumption associated with generating a density projection plot along the z-axis, which results in an overall $81\%$ increase in memory usage compared to the baseline case without \libyt.
}

\subsection{Limitations} \label{subsec:limitations}

{\libyt\ currently supports only patch-based AMR grid data structures.
This reflects the scope of the current implementation rather than a fundamental limitation.
The core steps: binding simulation data to Python, mapping data to \yt, and communication methods, can be adapted to other data layouts, which we plan to explore in future work.
}

{Beyond this limitation, other constraints of} \libyt\ are closely related to the mechanisms used to achieve in situ analysis for running arbitrary Python codes and supporting Python prompt-like entry points, allowing users to input and receive feedback instantly under MPI parallel computing.
They mainly arise from these two reasons: the first one is the data redistribution process (\sref{subsec:parallelism}), which requires every MPI process to participate; and the second one is the inputting of Python codes, waiting for the Python interpreter outputs, and the collecting of outputs in every MPI process cycle when supporting the Python shell (\sref{subsubsec:python-prompt-and-libyt-commands}).

We utilize collective one-sided MPI operations to perform data redistribution, which requires every MPI process to participate, thereby eliminating the need for point-to-point handshaking communication between nodes.
{While this collective approach provides an abstraction that allows any process to access data it does not own, it introduces global synchronization overhead that may affect scalability, as shown in \sref{subsubsec:weak-strong-scaling-libyt}.}

{The in situ analysis process} inherits the parallelism feature in \yt, including decomposing and distributing jobs to MPI processes,
\texttt{yt\_libyt} (\yt\ frontend for \libyt, see \sref{subsec:connecting-libyt-to-yt}), which {only passes simulation data to} \yt\ and is not involved in job decomposition, cannot determine in advance whether each MPI process already has the required data for a given \yt\ operation.
Consequently, \texttt{yt\_libyt} performs an initial synchronization step to determine the data needed by each MPI process, making all {data-reading} operations collective.
Thus, even if the \yt\ operation satisfies data locality (i.e., every MPI process only needs the data it already owns), 
\texttt{yt\_libyt} still requires all MPI processes to enter the data-reading phase.
As a result, every MPI process must run the same \yt\ operation in an inline Python script whenever the operation {reads data}; otherwise, the program will hang and fail.
Fortunately, most \yt\ operations read data synchronously. 

Some \yt\ functions may violate the above condition.
For example, for a figure annotated with streamlines using \texttt{annotate\_streamlines}, 
\yt\ needs to perform {an additional data-reading operation} when calling the \texttt{save} method to save this figure.
This is straightforward in post-processing since all MPI processes can access data on disk independently, so only one process needs to call \texttt{save}.
By contrast, for \libyt, all MPI processes must call \texttt{save} to avoid hanging due to {the collective operation in the data-reading phase}. 
Several annotation functions, such as \texttt{annotate\_quiver}, \texttt{annotate\_cquiver}, \texttt{annotate\_velocity}, \texttt{annotate\_magnetic\_field}, and \texttt{annotate\_particles}, also have the same limitation. 
However, saving the same figure redundantly is not a serious problem as long as it does not corrupt an output file, since this operation is not time-consuming.
Volume rendering in \yt\ {includes an asymmetric data-reading phase during figure output}.
{If the number of MPI processes is odd, some processes may not participate in this phase.}
A compromised solution is thus to launch an even number of processes and {ensure that each process invokes} \texttt{save}.

{The above limitations---reduced scalability at large processor counts and the need for all processes to enter the data-reading phase---}are mainly caused by {the collective data-reading} mechanism currently implemented in \libyt.
{These issues could be addressed either by adopting a non-collective data-reading approach using a hierarchical architecture, in which a central coordinator dispatches tasks instead of using a flat structure}, or by revising the way \yt\ decomposes and distributes tasks.
We {plan to} address these limitations in the future.

To support an interactive Python shell, where users can enter Python code and receive immediate feedback while coordinating \textit{N} MPI processes, \libyt\ redirects the output from the Python interpreters in all processes, waits for their outputs, aggregates them, and prints the combined output.
It isolates the Python interpreter and restricts communication to Python code or text, so only strings pass into and out of the wrapper.
While this method successfully manages every MPI process, it has a defect in that it cannot directly access the Python interpreter and is unaware of its current state.
\libyt\ can only send the code and await the result until it is returned.
Consequently, \libyt\ cannot handle certain operations, such as \texttt{KeyboardInterrupt}, when interrupting the kernel using the Jupyter Notebook frontend.
Moreover, since most Jupyter Notebook extensions display results in formats (e.g., HTML, SVG, JavaScript, etc) other than text, users cannot use them because \libyt\ only returns and displays everything in plain text.
Another caveat is that \libyt\ slows down as output strings grow.
We will address these flaws by introducing mechanisms for detecting and displaying different formats and by improving the efficiency and flexibility of the output-gathering routine across all MPI processes in future development.

\subsection{Comparing Libyt to Traditional Methods} \label{subsec:comparing-to-traditional-methods}

Both \libyt\ and post-processing allow Python to access the data.
The former uses the \libyt\ Python Module (\sref{subsec:embedding-python-in-c++-application}) to retrieve data from distributed memory across multiple computing nodes, and the latter directly reads data from disk.
During \libyt\ in situ analysis, the runtime switches from the simulation code to Python interpreters,
and all simulation processes pause until the Python routines complete.
Each approach has its own advantages and limitations, and the appropriate choice depends on the specific scientific task or computational requirements.

\libyt\ has features for exposing entry points (interactive Python prompt in \sref{subsec:interactive-mode} and Jupyter Notebook in \sref{subsec:support-jupyter-notebook}) for users to interact with the simulation data.
During the interactive exploration of the data stage, the simulation halts until users have completed their tasks.
Preserving the whole simulation state at a timestamp is analogous to the restart mechanism in post-processing, 
where the simulation dumps all the necessary data to disk so that the state can be reproduced later.
The interactive feature in \libyt\ is handy when developing and scaling up simulations.
The restart method is more suitable when we want to analyze a dataset extensively or conduct an undirected exploration.

A typical hindrance when developing large-scale simulations is tracking subtle logistic errors in simulation algorithms that do not cause the program to crash in a high-performance computing cluster.
With \libyt's interactive feature, we can call the API at any potentially problematic location and investigate the simulation data using Python syntax.
A robust Python environment enables developers to explore the live simulation data, generate diagnostic plots, identify issues as they occur, and monitor bugs step-by-step under an MPI parallel computing setup.
The back-communication in \libyt\ further enables users to probe the underlying user-defined functions.
This would be expensive and complicated if we were to do it the traditional way.
We would need to store a series of data snapshots before examining them.
The traditional post-processing method is preferred when the data snapshot at a specific step needs to be examined intensively or when the scientific goals are not yet well defined.

Simulations involving heavy computational tasks utilize efficient programming languages such as C and C++. 
In contrast, many data analysis and visualization tools are developed in Python, or their user interfaces are exposed in Python syntax.
\libyt\ effectively combines the computational efficiency of C/C++ with the extensive capabilities and flexibilities of the Python ecosystem by calling Python functions during simulation runtime.
It provides a way to couple existing C/C++ and Python codes without reinventing the wheel.
\libyt\ can integrate that data analysis routine into the simulation processes when users have identified the quantities they wish to examine.
The interchangeability of post-processing and in situ Python scripts simplifies this integration.
Furthermore, instead of storing a series of data snapshots on disk, we can regard simulations, their initial parameters, and the inline Python script as a complete state for reproducing the data analysis process and the time series simulation results.

{Using \libyt\ to perform the same data analysis does not incur additional time compared to post-processing under the same computing resources when the processor count is moderate (up to approximately one hundred), as shown in \sref{subsubsec:time-compare-to-postprocessing}.}
Instead of storing a data snapshot and then analyzing it separately, like in post-processing,
\libyt\ makes the analysis workflow part of the simulation routine without increasing the overall time cost.
{However, at larger processor counts, the collective data-reading and redistribution mechanisms introduce communication overhead that can limit scalability (\sref{subsubsec:weak-strong-scaling-libyt}). }
{A further limitation emerges from holding the simulation's, \libyt's, and Python's runtime data concurrently in memory within a single process (as shown in \sref{subsec:memory-performance}), whereas post-processing typically incurs memory usage only for the Python runtime.}

In summary, \libyt\ provides a complementary alternative to traditional post-processing workflows, enabling Python-based data access and analysis directly within the simulation runtime.
The interchangeability of script design allows users to switch seamlessly between in situ inspection and conventional post-processing, depending on the use case.

\subsection{Linking Libyt to Other Libraries} \label{subsec:linking-to-other-tools}

Similar adapters can be created for other Python packages that utilize \texttt{mpi4py} or other MPI-equivalent Python implementations in parallel computing, allowing their post-processing scripts to be directly applied in our in situ analysis workflow.
Users can develop their own Python scripts, understanding that the simulation data are accessible through the \libyt\ Python Module.

\libyt\ also enables high-frequency data access for AI and machine learning workflows by providing a lightweight mechanism to stream simulation data directly into Python during runtime.
Because data can be retrieved without writing the intermediate snapshots to disk, AI and machine learning models can be updated, trained, or evaluated at fine temporal resolutions that would otherwise be extremely expensive with traditional post-processing methods.
This capability is particularly valuable for tasks such as adaptive modeling, dynamic feature extraction, and real-time anomaly detection, where rapid, repeated access to evolving simulation states is essential.
In addition, \libyt's back-communication interface provides a pathway for AI and machine learning models to write back into the simulation, enabling tightly coupled hybrid simulations that combine physics-based solvers with learned components, such as replacing time-consuming simulations with surrogate models.
We plan to demonstrate these capabilities further in future work.

Besides the extensibility in Python, we may link \libyt\ to other in situ analysis frameworks by creating an adapter between \libyt's interface and that of other frameworks.
{In addition, because \libyt\ is implemented as a C library, it can be interfaced with other programming languages (e.g., Fortran), and we plan to investigate such support in future development.}

\section{Conclusions} \label{sec:conclusions}

Data visualization and analysis are increasingly becoming bottlenecks in exascale simulations because of the {massive dataset, the disk reading and writing speed bottleneck,} and limited hard disk space. 
Several in situ analysis tools, which bypass the intermediate step of writing data to a hard disk and analyze the data directly in memory, have been introduced to address this issue. 
However, we still find the need to have a library that is flexible enough to run our everyday-used post-processing Python scripts using an already set up Python environment. 

This paper presents \libyt, an in situ library that seamlessly integrates the Python data analysis process into the simulation workflow and provides entry points like Jupyter Notebook and Python prompt for users to interact with the in-memory simulation data under MPI parallel computing.
The analysis becomes part of the routine, such that the simulation processes stop indefinitely and only resume after Python has done its job.
\libyt\ enables back-communication of simulation information requested by Python.
By integrating the Python data analysis workflow into simulations,
we merge repetitive jobs and save tons of disk space by only storing the processed data,
which would previously require embarrassingly parallelizing every data snapshot at every time step in traditional methods.
The entry points enable users to interact with distributed simulation data in-memory.
\libyt\ is free and open-source software with no non-free dependencies maintained on GitHub.
It provides interfaces for simulation to bind data and user-defined C functions to Python, enabling Python to read in-memory simulation data or request the simulation to generate data on demand. 

\libyt\ focuses on using \yt\ as its core analytic method.
A frontend is created to enable the interchangeability of post-processing and in situ Python scripts. 
Converting them is a two-line change.
Users can determine which Python environment to use at compilation and reuse the Python script, 
which enables us to borrow the existing Python ecosystem and lowers the barrier to adapting to this new tool. 
We demonstrate some production runs in \gamer\ and \enzo\ simulations and display how \libyt\ successfully addresses the problem of insufficient hard disk space to store all data snapshots by integrating the Python data analysis workflow into the simulation and storing only the necessary outputs and data.
This enables us to closely capture and monitor events within the simulation domain, as well as track the exact changes in the governing equations.
These lower the costs for running extensive simulations on high-performance computing platforms 
and enable us to conduct even more extensive simulations.
The time cost for \libyt\ to run the same Python workflow is faster than the post-processing method.
There is no additional time penalty.
There is a caveat in \libyt\ regarding the simulation, \libyt, and Python runtime data being kept in a single process, which may cause the requested memory to exceed the computing node's capacity.

\libyt\ bridges Python and simulations bidirectionally.
The reusability of Python scripts and simulations that are grounded in physical equations simplifies the development process.
It also provides an interactive Python prompt and Jupyter Notebook access point for users to explore the data.
These features blur the line between simulations and Python.
On a broader picture, we may use \libyt\ to integrate Python machine learning and AI frameworks into the simulations.
C/C++ simulations can benefit from Python's productivity and robust libraries, 
and Python can take advantage of C/C++'s fast and efficient features in computation-intensive tasks.


\begin{acknowledgments}
We gratefully acknowledge the support of the School of Information Sciences at the University of Illinois Urbana-Champaign for supporting Shin-Rong's research.
Hsi-Yu acknowledges funding support from the Yushan Young Scholar Award No. NTU-111V1201-5, sponsored by the Ministry of Education (MoE) of Taiwan.
This publication is also supported in part by the Gordon and Betty Moore Foundation's Data-Driven Discovery Initiative through Grants GBMF9123 and GBMF4561 to Matthew Turk. 
This work was also partially supported by the National Science Foundation through grants ACI-1339624 and ACI-1663914.
This research is partially supported by the National Science and Technology Council (NSTC) of Taiwan under Grants No. NSTC 111-2628-M-002-005-MY4 and No. NSTC 108-2112-M-002-023-MY3, and by the NTU Academic Research-Career Development Project under Grant No. NTU-CDP-113L7729.
We thank the National Center for High-performance Computing (NCHC) for providing computational and storage resources.
We are also grateful to He-Feng Hsieh for conducting the core-collapse supernova simulations and Hsinhao Huang for performing the isolated dwarf galaxies and fuzzy dark matter simulations with \gamer.
We also thank the National Center for Supercomputing Applications (NCSA) and the Joint Laboratory for Extreme-Scale Computing (JLESC) for hosting workshops and providing numerous inspiring talks and learning opportunities that helped shape this work.
We further acknowledge the Project Jupyter, upon which several components of this work are built. 
Finally, we are grateful for the support of the yt project community for their support and kindness.
\end{acknowledgments}

\vspace{5mm}

\software{
{\libyt\ \citep{libyt-zenodo}},
Python \citep{python3},
\yt\ \citep{yt},
\texttt{mpi4py} \citep{mpi4py},
NumPy \citep{numpy},
Matplotlib \citep{matplotlib},
{MPI \citep{mpi31}},
\gamer\ \citep{gamer-2},
\enzo\ \citep{enzo},
\texttt{valgrind} \citep{valgrind}
}

\appendix

\section{Memory Profile} \label{appendix:memory-profile}

\texttt{valgrind} samples the memory profile of the heap and the stack memory throughout a period using \texttt{massif} with \texttt{--stack=yes}.
To obtain a detailed snapshot when the program reaches a specific line of code, we need to insert the following line at that location to store the profile in \texttt{filename}.
We use this to closely track the memory profile of \libyt\ before and after each step.
We also hook this under \libyt\ Python Module method to dump a memory profile when reaching a specific piece of Python code.
\\
\\
{\small
\begin{tabular*}{\columnwidth}{l}
    \toprule
    \texttt{VALGRIND\_MONITOR\_COMMAND("detailed\_snapshot filename");} \\
    \hline
\end{tabular*}
}

To run \texttt{valgrind} under MPI parallel computation and track Python's memory consumption, we need to export the following:
\\
\\
{\small
\begin{tabular*}{\columnwidth}{l}
    \toprule
    \texttt{export LDPRELOAD=\$VALGRIND\_PREFIX/lib/valgrind/libmpiwrap-amd64-linux.so} \\
    \texttt{export PYTHONMALLOC=malloc} \\
    \hline
\end{tabular*}
}

\section{Memory Profile Python Scripts} \label{appendix:memory-profile-python-scripts}

The Python scripts are loaded in the separate memory performance tests, and \texttt{yt\_inline} Python is called.
We hook the \texttt{valgrind} command under \libyt\ Python Module (see Appendix \ref{appendix:memory-profile}), which makes the following code to dump a detailed memory snapshot.
\\
\\
{\small
\begin{tabular*}{\columnwidth}{l}
    \toprule
    \texttt{libyt.dump\_valgrind\_detailed\_snapshot(f"{profile\_filename}")} \\
    \hline
\end{tabular*}
}

In the tested Python scripts, we also extract code out of \texttt{yt\_inline} Python function that only needs to be initialized once, to mimic the actual Python code in a production run.

\subsection{Empty Python Function}

The Python script runs empty-function memory profiling. 
The function is not empty, due to the memory profile sampling. 
The variables in the function are negligible.
\\
\\
{\small
\begin{tabular*}{\columnwidth}{l}
    \toprule
    \texttt{from mpi4py import MPI} \\
    \texttt{import libyt} \\
    \\
    \texttt{my\_rank = MPI.COMM\_WORLD.Get\_rank()} \\
    \texttt{step = 0} \\
    \\
    \texttt{def yt\_inline():} \\
    \texttt{\quad \quad global step} \\
    \texttt{\quad \quad libyt.dump\_valgrind\_detailed\_snapshot(f"MPI\{my\_rank\}PythonFunc\{step\}")} \\
    \texttt{\quad \quad step += 1} \\
    \hline
\end{tabular*}
}

\subsection{Generating Derived Field Python Function}

The Python script runs derived field memory profiling. 
The memory profile is sampled after generating $5,000$ grids. 
Each generated grid contains $8^3$ cells with data type \texttt{float} and is filled with ones.
\\
\\
{\small
\begin{tabular*}{\columnwidth}{l}
    \toprule
    \texttt{from mpi4py import MPI} \\
    \texttt{import libyt} \\
    \\
    \texttt{my\_rank = MPI.COMM\_WORLD.Get\_rank()} \\
    \texttt{my\_size = MPI.COMM\_WORLD.Get\_size()} \\
    \texttt{start = my\_rank * int(2000000 / my\_size)} \\
    \texttt{num\_to\_generate = 5000} \\
    \texttt{step = 0} \\
    \\
    \texttt{def yt\_inline():} \\
    \texttt{\quad \quad global step} \\
    \texttt{\quad \quad data\_list = []} \\
    \texttt{\quad \quad for i in range(start, start + num\_to\_generate):} \\
    \texttt{\quad \quad \quad \quad data = libyt.derived\_func(i, "DerivedOnes")} \\
    \texttt{\quad \quad \quad \quad data\_list.append(data)} \\
    \texttt{\quad \quad libyt.dump\_valgrind\_detailed\_snapshot(f"MPI\{my\_rank\}PythonFunc\{step\}")} \\
    \texttt{\quad \quad step += 1} \\
    \hline
\end{tabular*}
}

\subsection{Redistributing Data Python Function}

The Python script runs the data redistribution memory profile.
The memory profile is sampled after finishing the redistribution of $20,000$ grids. 
There are four MPI processes.
Each MPI process prepares $5,000$ grids and fetches the other $15,000$ grids from the other three MPI processes.
Each grid contains $8^3$ cells with data type \texttt{float}.
\\
\\
{\small
\begin{tabular*}{\columnwidth}{l}
    \toprule
    \texttt{from mpi4py import MPI} \\
    \texttt{import libyt} \\
    \\
    \texttt{myrank = MPI.COMM\_WORLD.Get\_rank()} \\
    \texttt{mysize = MPI.COMM\_WORLD.Get\_size()} \\
    \texttt{num\_grids = 2000000} \\
    \texttt{num\_local = int(2000000 / mysize)} \\
    \texttt{start\_idx = myrank * num\_local} \\
    \texttt{prepare\_len = 5000} \\
    \texttt{prepare\_gids = list(range(start\_idx, start\_idx + prepare\_len))} \\
    \texttt{fetch\_gids, fetch\_rank = [], []} \\
    \texttt{for r in range(mysize):} \\
    \texttt{\quad \quad if r != myrank:} \\
    \texttt{\quad \quad \quad \quad fetch\_gids += list(range(r * num\_local, r * num\_local + prepare\_len))} \\
    \texttt{\quad \quad \quad \quad fetch\_rank += [r] * prepare\_len} \\
    \texttt{step = 0} \\
    \\
    \texttt{def yt\_inline():} \\
    \texttt{\quad \quad global step} \\
    \texttt{\quad \quad data = libyt.get\_field\_remote(["CCTwos".encode(encoding="UTF-8", errors="strict")], 1, } \\
    \texttt{\quad \quad \quad \quad \quad \quad \quad \quad \quad \quad \quad \quad \quad \quad \quad \quad \quad prepare\_gids, len(prepare\_gids), } \\
    \texttt{\quad \quad \quad \quad \quad \quad \quad \quad \quad \quad \quad \quad \quad \quad \quad \quad \quad fetch\_gids, fetch\_rank, len(fetch\_gids))} \\
    \texttt{\quad \quad libyt.dump\_valgrind\_detailed\_snapshot(f"MPI\{myrank\}PythonFunc\{step\}")} \\
    \texttt{\quad \quad step += 1} \\
    \hline
\end{tabular*}
}

\subsection{yt Operation Print Data Snapshot States}

The Python script runs the \yt\ operation \texttt{print\_stats} with the memory profile. 
The operation retrieves basic information about the dataset, such as the number of grids and the number of cells per level.
\\
\\
{\small
\begin{tabular*}{\columnwidth}{l}
    \toprule
    \texttt{from mpi4py import MPI} \\
    \texttt{import libyt} \\
    \texttt{import yt\_libyt} \\
    \texttt{import yt} \\
    \\
    \texttt{myrank = MPI.COMM\_WORLD.Get\_rank()} \\
    \texttt{mysize = MPI.COMM\_WORLD.Get\_size()} \\
    \texttt{yt.enable\_parallelism()} \\
    \texttt{yt.set\_log\_level(50)} \\
    \texttt{step = 0} \\
    \\
    \texttt{def yt\_inline():} \\
    \texttt{\quad \quad global step} \\
    \texttt{\quad \quad ds = yt\_libyt.libytDataset()} \\
    \texttt{\quad \quad ds.print\_stats()} \\
    \texttt{\quad \quad libyt.dump\_valgrind\_detailed\_snapshot(f"MPI\{my\_rank\}PythonFunc\{step\}")} \\
    \texttt{\quad \quad step += 1} \\
    \hline
\end{tabular*}
}

\bibliography{libyt}{}
\bibliographystyle{aasjournalv7}


\end{CJK*}

\end{document}